\documentclass[pdflatex,sn-mathphys-num,twocolumn]{sn-jnl}
\usepackage{graphicx}%
\usepackage{multirow}%
\usepackage{amsmath,amssymb,amsfonts}%
\usepackage{amsthm}%
\usepackage{mathrsfs}%
\usepackage[title]{appendix}%
\usepackage{xcolor}%
\usepackage{textcomp}%
\usepackage{manyfoot}%
\usepackage{booktabs}%
\usepackage{algorithm}%
\usepackage{algorithmicx}%
\usepackage{algpseudocode}%
\usepackage{listings}%
\usepackage{tikz}%
\usepackage{placeins}
\usepackage{dblfloatfix}
\usepackage{subcaption}%
\usepackage{algpseudocode}%
\usepackage{microtype}%
\theoremstyle{thmstyleone}%
\newcounter{mainfigurecount}
\theoremstyle{thmstyletwo}%
\usepackage{xspace}
\newcommand{\um}{\text{\textmu m}\xspace}
\newcommand{\microsec}{\text{\textmu s}\xspace}

\theoremstyle{thmstylethree}%

\allowdisplaybreaks

\usetikzlibrary{arrows.meta, positioning, calc}

\begin{document}
\title[Component-Level Inverse Design of Transmon Qubits]{Component-Level Inverse Design of Transmon Qubits Using Neural Networks}

\author*[1,2]{\fnm{Olivia} \sur{Seidel}}\email{olivias@fnal.gov}
\author[3]{\fnm{Firas} \sur{Abouzahr}}\email{firasabouzahr2030@u.northwestern.edu}
\author[4,5]{\fnm{Abhishek} \sur{Chakraborty}}\email{achakraborty@chapman.edu}
\author[6,7]{\fnm{Sadman Ahmed} \sur{Shanto}}\email{shanto@usc.edu}
\author[6,7]{\fnm{Saikat} \sur{Das}}\email{saikatda@usc.edu}
\author[1,3]{\fnm{Daniel} \sur{Baxter}}\email{dbaxter9@fnal.gov}
\author[2]{\fnm{Jonathan} \sur{Asaadi}}\email{jonathan.asaadi@uta.edu}
\author[10]{\fnm{Nicola} \sur{Pancotti}}\email{npancotti@nvidia.com}
\author[10]{\fnm{Haoyu} \sur{Yang}}\email{haoyuy@nvidia.com}
\author[10]{\fnm{Brucek} \sur{Khailany}}\email{bkhailany@nvidia.com}
\author[1]{\fnm{Sara} \sur{Sussman}}\email{sarafs@fnal.gov}
\author[1,3]{\fnm{Enectali} \sur{Figueroa-Feliciano}}\email{enectali@northwestern.edu}
\author[6,7,8,9]{\fnm{Eli M} \sur{Levenson-Falk}}\email{elevenso@usc.edu}
\author[10]{\fnm{Taylor L.} \sur{Patti}}\email{tpatti@nvidia.com}

\affil*[1]{\orgname{Fermi National Accelerator Laboratory},
  \orgaddress{\street{PO Box 500}, \city{Batavia}, \postcode{60510}, \state{IL}, \country{USA}}}

\affil[2]{\orgdiv{Department of Physics},
  \orgname{University of Texas at Arlington},
  \orgaddress{\street{502 Yates Street}, \city{Arlington}, \postcode{76019}, \state{TX}, \country{USA}}}

\affil[3]{\orgdiv{Department of Physics and Astronomy},
  \orgname{Northwestern University},
  \orgaddress{\street{2145 Sheridan Road}, \city{Evanston}, \postcode{60208}, \state{IL}, \country{USA}}}

\affil[4]{\orgdiv{Institute for Quantum Studies},
  \orgname{Chapman University},
  \orgaddress{\street{One University Drive}, \city{Orange}, \postcode{92866}, \state{CA}, \country{USA}}}

\affil[5]{\orgdiv{Department of Physics and Astronomy},
  \orgname{University of Rochester},
  \orgaddress{\street{500 Wilson Boulevard}, \city{Rochester}, \postcode{14627}, \state{NY}, \country{USA}}}

\affil[6]{\orgdiv{Center for Quantum Information Science and Technology},
  \orgname{University of Southern California},
  \orgaddress{\street{925 Bloom Walk}, \city{Los Angeles}, \postcode{90089}, \state{CA}, \country{USA}}}

\affil[7]{\orgdiv{Department of Physics \& Astronomy},
  \orgname{University of Southern California},
  \orgaddress{\street{825 Bloom Walk}, \city{Los Angeles}, \postcode{90089}, \state{CA}, \country{USA}}}

\affil[8]{\orgdiv{Ming Hsieh Department of Electrical \& Computer Engineering},
  \orgname{University of Southern California},
  \orgaddress{\street{3740 McClintock Avenue}, \city{Los Angeles}, \postcode{90089}, \state{CA}, \country{USA}}}

\affil[9]{\orgname{Quantum Elements, Inc},
  \orgaddress{\city{Thousand Oaks}, \state{CA}, \country{USA}}}

\affil[10]{\orgdiv{Research},
  \orgname{NVIDIA Corporation},
  \orgaddress{\street{2788 San Tomas Expressway}, \city{Santa Clara}, \postcode{95051}, \state{CA}, \country{USA}}}

\abstract{Designing a superconducting qubit to realize specific Hamiltonian parameters typically requires iterating through a time and compute-intensive forward loop in which the designer chooses a layout geometry, simulates it, extracts circuit parameters such as capacitances, and refines the geometry. We study the inverse version of this task using a neural-network workflow that maps target Hamiltonian parameters directly to component-level layout parameters, which we subsequently demonstrate on a planar transmon layout. During training, we pair the inverse model with a frozen forward surrogate model and evaluate the loss in Hamiltonian space rather than in layout-parameter space. In validation against a conventional EM solver, 97\% of generated designs produce usable geometries, and the inverse-plus-surrogate pipeline reaches mean percent errors of 0.73\% for qubit frequency and 1.58\% for anharmonicity, comparable to or below the fabrication and simulation-to-measurement uncertainty expected for academic-process transmon devices of this type. A single pipeline query takes $\sim$60\,ms on CPU, versus $\sim$2\,min for a conventional EM capacitance extraction on the same hardware, a speedup of approximately 2{,}000$\times$. Batching minimizes the AI model inference overhead, reducing the runtime to 3.1\,\microsec per sample on CPU and 2.6\,\microsec per sample on GPU at a batch size of 2048, resulting in speedups of $3.9\times10^{7}$ and $4.6\times10^{7}$, respectively, relative to a single conventional CPU EM extraction. Our results indicate that component-level inverse design usefully extends and complements conventional EM simulation, including for small datasets on the order of 1{,}000 samples.}

\maketitle

\section{Introduction}\label{sec:introduction}

Systematically designing superconducting qubits remains challenging, as designers still organize much of the design process as a forward loop. That is, the designer sketches a layout in a tool such as Quantum Metal (formerly known as Qiskit Metal) \cite{bib9, Wang2021QiskitMetal}, renders it for finite-element electromagnetic (EM) simulation with a conventional EM solver, extracts capacitance, impedance, and/or mode properties, computes the corresponding Hamiltonian parameters using circuit-quantization workflows \cite{Blais2004CircuitQED,Nigg2012BlackBox,bib12,bib13}, and then intuitively adjusts the geometry based on this result \cite{bib7, bib10}. Such a loop is not only heuristic and non-systematic, but it is also costly in both compute and human time. Moreover, the reliance on human intuition and expensive simulations is particularly problematic due to the fact that small geometric changes can have large and even counterintuitive effects on the effective circuit parameters. Although qubit modalities such as the transmon are well-characterized \cite{bib1}, general superconducting quantum processing unit (QPU)-design remains an engineering challenge that requires balancing circuit parameters, coherence, control, readout, materials, packaging, and fabrication constraints \cite{Krantz2019QuantumEngineer,Kjaergaard2020SuperconductingQubits,Blais2021CircuitQED,Devoret2013SuperconductingCircuits}. Because no direct map from a target Hamiltonian to a QPU layout is available, designing a manufacturable QPU that achieves a target Hamiltonian remains a slow, expert-driven cycle of repeated simulation.

Various methodologies have been explored to address computationally intensive problems in other fields by mapping target properties to the physical parameters that would realize them. These methods range from traditional statistical sampling to modern techniques based on generative AI. Traditionally, Bayesian approaches such as Markov-Chain Monte Carlo (MCMC) have been widely used for inverse problems in semiconductor and fluid dynamics design applications \cite{Stuart_2010, Cotter_2009, taghizadeh2024bayesianinversionidentificationdoping}. By leveraging random walk-based flows, an MCMC can identify satisfactory solutions, but it incurs significant computational overhead, particularly when the design space requires solving complex partial differential equations \cite{mcmc,fno,yang2025deep}.
While new generative AI models now allow for the direct learning of inverse operators, mapping property space to design space \cite{huang2024data,long2024invertible}, these models are often data-demanding and struggle with ill-posed problems. In the context of superconducting device geometry, a more direct route involves learning an inverse map from target Hamiltonian parameters to physical geometry. Similar strategies in nanophotonics demonstrate that AI can serve as both a fast surrogate for solvers and an inverse predictor for device geometries \cite{bib2,bib3,bib4}.

A primary complication of this approach remains the ill-posed nature of the problem. Multiple physical geometries can yield nearly identical Hamiltonian responses, and so a model that uses only a geometry-matching objective may ``average'' incompatible solutions, resulting in designs that are not physically viable \cite{bib3}. To circumvent this, recent literature suggests coupling an inverse network to a pre-trained forward surrogate, training the model based on predicted responses rather than geometry labels \cite{bib3,yang2025deep}. We adopt this tandem architecture here, applying it specifically to the automated design of transmons in Quantum Metal.

Similar ideas are starting to appear in superconducting-circuit workflows. Automated and optimization-driven design has been explored for superconducting circuit architectures and couplers~\cite{Menke2021Automated}, while more recent work has begun to use machine-learning models for parameter design, layout, and optimization~\cite{NugrahaShao2023,bib5,bib6,Yaker2026InverseSRF}. Prior work has shown that machine learning can be used for both forward and inverse modeling of individual Quantum Metal transmon layouts, using simulated Quantum Metal, finite-element EM, and lumped-oscillator-model data to connect geometry parameters with qubit properties such as qubit frequency~\cite{NugrahaShao2023}. Concurrent work has also applied neural networks to inverse design of superconducting radio-frequency cavities and transmons for bosonic quantum computation~\cite{Yaker2026InverseSRF}, targeting 3D cavity-transmon systems rather than the planar, layout-parameterized geometries we consider.

In this manuscript, we present a Hamiltonian-targeted inverse-design workflow comprised of a parameterized layout, a learned forward surrogate, and a tandem inverse network. While our method carries over to any design tool that supports parameterized layout generation, in this initial application, we test the workflow on a single cross-claw transmon design, where the model maps target qubit frequency $f_q$ and anharmonicity $\alpha$ to a set of design geometry parameters. We obtain training data from the SQuADDS open-source database of device designs and simulations \cite{bib7}. Although the available cross-claw transmon dataset has less than 2{,}000 samples, we show that it is large enough to train both the surrogate and the inverse model.

In this manuscript, we make the following contributions:
\begin{itemize}
\item An efficient forward surrogate multilayer perceptron (MLP) model for rapid inference that uses Quantum Metal geometry parameters for high-accuracy prediction of two key Hamiltonian parameters, qubit frequency $f_q$ and anharmonicity $\alpha$ (average error of 0.136\% and 0.293\%, respectively).
\item An inverse-design workflow demonstrated for cross-claw transmon Quantum Metal parameters, in which the forward surrogate is used to train an inverse MLP that maps target $f_q$ and $\alpha$ to a corresponding design geometry with high accuracy (median error of 0.53\% and 1.14\%, respectively).
\item A small-data scaling analysis and a nearest-neighbor stress test of the surrogate model.
\item Conventional EM-solver validation of the trained pipeline on SQuADDS cross-claw transmon samples.
\item Runtime benchmarking comparing conventional EM capacitance extraction with single-design and batched neural-network evaluation, demonstrating orders-of-magnitude speedup for candidate design generation and validation.
\end{itemize}

The rest of this manuscript proceeds as follows: Section~\ref{sec:theory} defines the target quantities and the physics-based mappings. Section~\ref{sec:data} describes data generation and pre-processing. Section~\ref{sec:methods} gives the model and training procedure. Section~\ref{sec:results} reports the cross-claw transmon results and small-data analysis. Finally, Sec.~\ref{sec:discussion} summarizes our results, their impacts, and the future research trajectories that they motivate.

\begin{figure*}[!tp] 
\centering
\captionsetup[subfigure]{labelformat=parens,labelsep=none}
\begin{subfigure}[t]{0.38\textwidth}
    \centering
    \includegraphics[width=\linewidth]{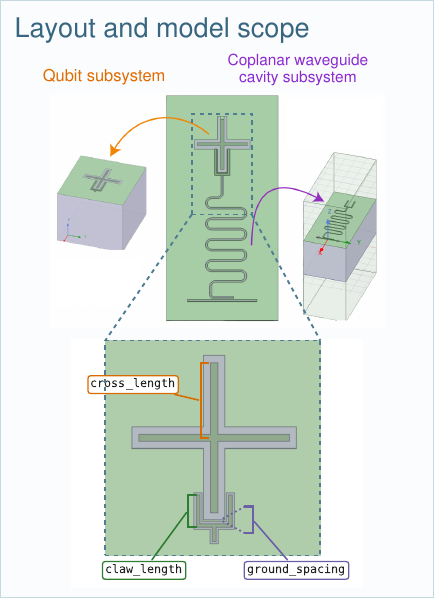}
    \caption{}
    \label{fig:overview-layout}
\end{subfigure}
\hfill
\begin{subfigure}[t]{0.53\textwidth}
    \centering
    \includegraphics[width=\linewidth]{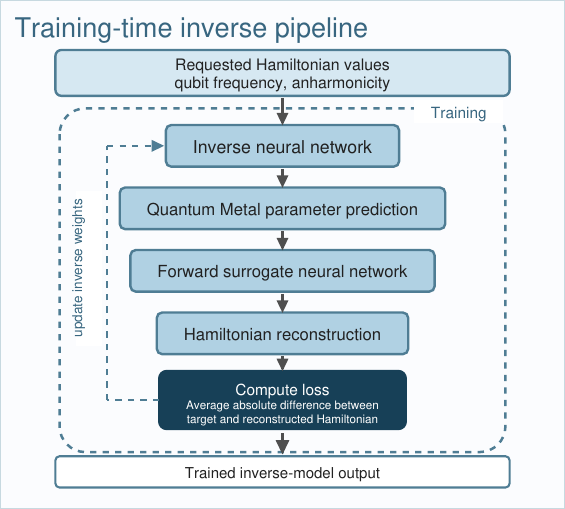}
    \caption{}
    \label{fig:overview-training}
\end{subfigure}

\vspace{0.3em}

\begin{subfigure}[t]{0.82\textwidth}
    \centering
    \includegraphics[width=\linewidth]{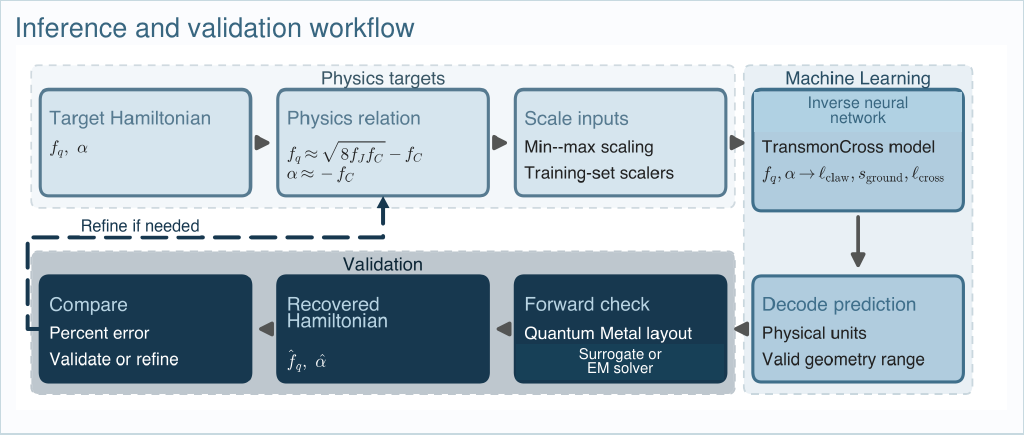}
    \caption{}
    \label{fig:overview-inference}
\end{subfigure}

\caption{Overview of the cross-claw transmon inverse-design workflow. (a) Layout context showing the cross-shaped transmon qubit and the claw-shaped coupling capacitor to a coplanar waveguide cavity, with a zoomed-in view of the transmon cross-claw section annotating the three geometry variables predicted by the inverse model: the cavity coupling claw length, the coupler ground spacing, and the transmon cross length. We show the cavity to clarify how this geometry is typically implemented, though in this work we focus only on the cross-claw subsystem. (b) Diagram of the training loop. The requested Hamiltonian values are mapped to geometry by the inverse neural network, passed through a frozen EM surrogate, and finally are compared with reconstructed Hamiltonian values. In this process, only the inverse-model weights are updated. (c) Inference and validation workflow. The requested Hamiltonian values are prepared, mapped to Quantum Metal geometry, decoded into a physical layout, and checked with a surrogate model or conventional EM capacitance solver before accepting or refining the design.}
\label{fig:overview}
\end{figure*}

\section{Background}\label{sec:theory}
This section defines the Hamiltonian targets, the physics that connects them to the device capacitances, and the geometry variables the inverse model predicts.

\subsection{Deriving Hamiltonian targets from the capacitance matrix}\label{subsec:physics-mapping}

In this work, the Hamiltonian targets are the qubit transition frequency $f_q$ and anharmonicity $\alpha$, which define the effective transmon Hamiltonian
\begin{equation}
\frac{\hat H_q}{h} = f_q \hat n_q + \frac{\alpha}{2}\hat n_q\left(\hat n_q-1\right)
\label{eq:transmon_hamiltonian}
\end{equation}

\noindent where $\hat n_q=\hat b^\dagger \hat b$ is the transmon excitation-number operator. Here, $\hat b$ and $\hat b^\dagger$ are the annihilation and creation operators for the transmon mode, $f_q = f_{01}$ is the frequency of the $0\!\to\!1$ transition, and $\alpha = f_{12}-f_{01} < 0$ for a transmon. Eq.~\eqref{eq:transmon_hamiltonian} is the standard Duffing-oscillator approximation to the transmon Hamiltonian, which is appropriate here because $f_q$ and $\alpha$ are exactly the two parameters it exposes.
In the transmon limit, the relations
\begin{equation}
\begin{gathered}
f_q \approx \frac{\sqrt{8E_JE_C}-E_C}{h},
\qquad
\alpha \approx -\frac{E_C}{h}
\end{gathered}
\label{eq:transmon-relations}
\end{equation}
show how the qubit Hamiltonian is set by the Josephson energy $E_J$ and charging energy $E_C$ \cite{bib1,Krantz2019QuantumEngineer,Blais2021CircuitQED}. The Josephson energy depends on the junction geometry and the fabrication process and is often not included in EM simulation except as a given parameter. The charging energy $E_C \equiv e^2/2 C_\Sigma$ depends on the total capacitance shunting the Josephson junction $C_\Sigma$. A finite-element electrostatic solver determines this capacitance, reporting a capacitance matrix for the structures in the geometry. SQuADDS includes the capacitance matrix, and the SQuADDS front end calculates Hamiltonian parameters based on the capacitance and $E_J$. Note that SQuADDS uses scqubits~\cite{bib13} to numerically calculate Hamiltonian parameters and does not rely on the analytical approximations above. In this work we hold $E_J$ fixed so that it gives an effective inductance of 10\,nH. Therefore, the inverse model is essentially learning to predict Quantum Metal geometry parameters that affect capacitance from target Hamiltonian parameters. With $E_J$ fixed, $f_q$ and $\alpha$ are not independent, because both are set by the single free quantity $E_C$, so targets lie on the one-dimensional curve in the $(f_q,\alpha)$ plane traced out by Eq.~\eqref{eq:transmon-relations}, and the inverse task is effectively a map from one Hamiltonian degree of freedom to three geometry parameters. All targets used in this work are drawn from the SQuADDS dataset and lie on this fixed-$E_J$ curve by construction. An arbitrary $(f_q,\alpha)$ pair off the curve is outside the training distribution, so user-specified targets should satisfy the fixed-$E_J$ relation rather than be chosen independently.

\subsection{Design variables and Hamiltonian targets}\label{subsec:variables-targets}
We use the Quantum Metal cross-claw transmon layout as the first test case for this workflow (Fig.~\ref{fig:overview-layout}). The device consists of a cross-shaped capacitor plate surrounded by a ground plane. A Josephson junction connects the cross and ground plane, forming the typical transmon circuit. The device also includes a ``claw'' that straddles one arm of the cross and serves as the other plate of a coupling capacitance. This cross-claw structure is commonly used to couple a transmon to a coplanar waveguide cavity as shown in the figure. This design is a useful controlled example because it has a small number of physically meaningful geometry parameters, while still producing the capacitance changes that alter the transmon Hamiltonian. In this study, we vary the coupling-claw length $\widehat{\ell}_{\mathrm{claw}}$, the width of the strip of ground plane between claw and cross (which Quantum Metal refers to as ``ground spacing'') $\widehat{s}_{\mathrm{ground}}$, and the length of the cross arms $\widehat{\ell}_{\mathrm{cross}}$, while keeping all other layout settings fixed, as shown in Fig.~\ref{fig:overview-layout}. This lets us isolate how changes in the cross-claw transmon geometry map to qubit frequency and anharmonicity without mixing in additional layout degrees of freedom. We choose this layout because it is simple enough to provide a controlled first demonstration of the inverse-design workflow, but still representative of the component-level design loop used in superconducting-qubit workflows.

In the SQuADDS-derived dataset used here, the geometry parameters span \(\ell_{\mathrm{claw}} \in [70\,\um,\, 400\,\um]\), \(s_{\mathrm{ground}} \in [4\,\um,\, 10\,\um]\), and \(\ell_{\mathrm{cross}} \in [90\,\um,\, 420\,\um]\). The junction inductance is fixed at 10\,nH, so different targets of \(f_q\) and \(\alpha\) are determined by the capacitance set by the transmon geometry.

\section{Data generation and preprocessing}\label{sec:data}

This section describes how we construct, filter, and prepare the cross-claw transmon dataset for training. Each sample links a set of Quantum Metal geometry parameters to the corresponding Hamiltonian quantities used as inverse-model targets. We first describe the SQuADDS-derived component dataset and the parameter fields retained for the inverse-design task, then summarize the EM-solver/scqubits extraction workflow that produces $f_q$ and $\alpha$, and finally describe the normalization procedure used for neural-network training.

\subsection{Cross-claw transmon component dataset and sampling strategy}\label{subsec:data-sampling}
We use the SQuADDS database~\cite{bib7} as the source for the cross-claw transmon component dataset and as the reference workflow for forward validation. Starting from the cross-claw transmon SQuADDS parameter set, one of various datasets in the database, we keep only the fields that vary within the dataset: the coupling-claw length, ground spacing, and transmon cross length. We hold all other SQuADDS layout and material settings fixed at their default values. 

The resulting cross-claw transmon dataset contains 1{,}934 existing SQuADDS samples. We do not generate additional random training samples for this split. Instead, we randomly partition the existing records into training, validation, and test subsets using a fixed random seed for reproducibility. Specifically, 70\% of the records are assigned to training, and the remaining 30\% are split evenly into validation and test sets, giving 1{,}353 training samples, 290 validation samples, and 291 test samples. The training set is used to update the neural network weights, the validation set is used during hyperparameter selection and early stopping, and the test set is not used during training or model selection, and is instead reserved for final evaluation on unseen SQuADDS records. Fig.~\ref{fig:sample-data-distribution} in Appendix~\ref{app:training-details} shows that this fixed split preserves the overall parameter coverage across the three Quantum Metal outputs defined in Section~\ref{subsec:variables-targets}.

Across the three splits, the dataset favors smaller claw lengths, samples ground spacing on a coarse grid, and distributes cross length more evenly. The similar train, validation, and test distributions indicate that the fixed 70/15/15 split preserves the overall distribution of the geometry parameters used for training and evaluation. Because ground spacing is sampled only on a coarse discrete grid, predictions between sampled values are less directly supported by the training data.

\subsection{Simulation pipeline}\label{subsec:simulation-pipeline}
We set up a pipeline for performing EM simulations to independently evaluate the layouts generated by our model. Each test sample starts from a Quantum Metal design template matching the form of entries in the public SQuADDS database \cite{bib7}, with the free geometry parameters set to what is predicted by the model. The resulting Hamiltonian parameters $f_q$ and $\alpha$ are obtained using Ansys Q3D, the solver originally used to generate the public SQuADDS database, together with the scqubits workflow described in Section~\ref{subsec:physics-mapping}. We use the same solver settings as those used originally to generate the SQuADDS data, and we have confirmed that we reproduce SQuADDS Hamiltonian parameters within 0.005\% for qubit frequency and 0.01\% for anharmonicity.

Each SQuADDS sample contains paired Quantum Metal geometry parameters and Hamiltonian quantities used for model training, while additional EM simulations are performed only to independently evaluate the model-generated designs. During these simulations, solver convergence is monitored through the relative error between adaptive passes. 

\subsection{Scaling and normalization}\label{subsec:scaling}

We apply min-max normalization \cite{sklearn_minmax_scaler} to both input and output Hamiltonian parameter vectors and Quantum Metal geometry vectors. The scalers are fitted using the training set only and then applied to the validation and test sets. This prevents information from the validation and test distributions from leaking into the preprocessing fit. Both the inverse MLP and the forward surrogate MLP are trained in scaled coordinates. At evaluation time, predicted quantities are inverse-transformed back into physical units before computing reported errors or passing candidate geometries to downstream solver-based evaluation.

For each Quantum Metal geometry parameter $y_j$, we apply min-max scaling using the training-set range,
\begin{equation}
\tilde{y}_{ij} = \frac{y_{ij}-y^{\mathrm{train}}_{j,\min}} {y^{\mathrm{train}}_{j,\max}-y^{\mathrm{train}}_{j,\min}}
\label{eq:minmax-geometry}
\end{equation}
\noindent where $i$ indexes the sample, $j$ indexes the geometry parameter, and $y^{\mathrm{train}}_{j,\min}$ and $y^{\mathrm{train}}_{j,\max}$ are the minimum and maximum values of parameter $j$ over the training set only. For the cross-claw transmon model, the scaled geometry vector is
\begin{equation}
\tilde{\mathbf{y}}_i = (\tilde{\ell}_{\mathrm{claw},i}, \tilde{s}_{\mathrm{ground},i}, \tilde{\ell}_{\mathrm{cross},i}).
\label{eq:scaled-geometry-vector}
\end{equation}
We apply the same training-set-only scaling procedure to the Hamiltonian parameter vector before neural-network training.

\FloatBarrier
\section{Models and training}\label{sec:methods}
We use multi-layer perceptrons (MLPs) for both the forward surrogate and inverse models because the inputs and outputs of both modeling tasks are low-dimensional numerical parameters, and MLPs can learn complex relationships between a small number of numerical inputs and outputs \cite{hornik1989multilayer}. The inverse model maps two Hamiltonian quantities (one effective degree of freedom at fixed $E_J$, as discussed in Section~\ref{subsec:physics-mapping}) to three Quantum Metal geometry parameters, while the surrogate maps the same three geometry parameters back to two Hamiltonian quantities. For the inverse model, we use a compact MLP because the dataset contains fewer than 2{,}000 samples, and larger networks can overfit sparsely sampled regions of the design space. We select the architecture through the hyperparameter sweep detailed below. Fully connected networks of comparable scale are also common in neural-network approaches to nanophotonic inverse design~\cite{bib2,bib3,bib4}.

We use the two networks in a tandem configuration. The map from Hamiltonian targets back to Quantum Metal parameters is non-unique, as different layouts can produce different capacitance matrices that yield nearly the same total shunt capacitance and therefore nearly identical Hamiltonian values, which we refer to as the ``one-to-many inverse problem''. If the inverse model is trained with only a direct geometry-regression loss, several valid designs compete for the same target and the network can average them into a poor design, the failure mode noted above. To avoid this, the forward surrogate is trained first to approximate the EM-solver/scqubits mapping from Quantum Metal geometry to Hamiltonian parameters. It is then frozen and placed after the inverse model, so that the inverse model is trained by comparing the recovered Hamiltonian values with the requested targets. This makes the training objective Hamiltonian-based rather than geometry-based, and using the surrogate greatly speeds up training, as each EM capacitance extraction takes minutes (Section~\ref{subsec:runtime}) and evaluating a design sweep therefore requires hours of solver time.

\begin{figure}[!t]
\centering
\includegraphics[width=\linewidth]{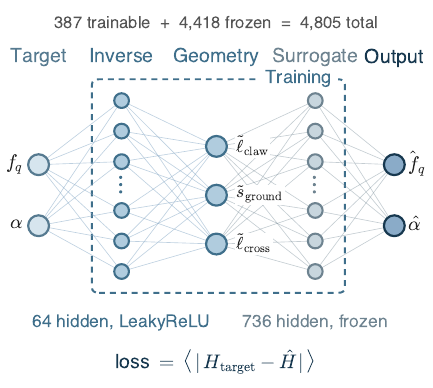}
\caption{Our inverse-plus-surrogate tandem architecture. The inverse MLP maps the target Hamiltonian parameters $(f_q,\alpha)$ to scaled Quantum Metal geometry $(\tilde{\ell}_{\mathrm{claw}},\tilde{s}_{\mathrm{ground}},\tilde{\ell}_{\mathrm{cross}})$ through one hidden layer of width 64 with LeakyReLU activations (387 trainable parameters). The frozen forward surrogate (736 hidden units, 4{,}418 non-trainable parameters) reconstructs the Hamiltonian $(\hat{f}_q,\hat{\alpha})$. Gradients of the mean absolute difference $\langle |H_{\mathrm{target}} - \hat{H}|\rangle$ flow back through the surrogate but update only the inverse-MLP weights.}
\label{fig:model-architecture}
\end{figure}

The inverse MLP has one hidden layer of width 64 with LeakyReLU activations (negative slope 0.01), a linear output layer, and 387 trainable parameters in total. It maps the two Hamiltonian targets $(f_q, \alpha)$ to the three scaled Quantum Metal parameters $\tilde{\ell}_{\mathrm{claw}}, \tilde{s}_{\mathrm{ground}}, \tilde{\ell}_{\mathrm{cross}}$. The frozen forward surrogate is a larger MLP with 736 units, the same LeakyReLU and linear output activations, and 4{,}418 parameters, trained on the same SQuADDS-derived dataset to map geometry back to $(f_q, \alpha)$, as shown in Fig.~\ref{fig:model-architecture}. During inverse-model training we update only the inverse-MLP weights. This training structure is analogous to the one Liu et al.\ used for nanophotonic inverse design \cite{bib3}, adapted here to Quantum Metal design variables.

The inverse model is trained on the loss
\[\mathcal{L}_{\mathrm{total}}=\mathcal{L}_{\mathrm{Ham}}+\lambda_{\mathrm{range}}\mathcal{L}_{\mathrm{range}},\]
where
\begin{equation}
\mathcal{L}_{\mathrm{Ham}}=\frac{1}{2N}\sum_{i=1}^{N}\left(\left|\tilde{f}_{q,i}-\hat{\tilde{f}}_{q,i}\right|+\left|\tilde{\alpha}_{i}-\hat{\tilde{\alpha}}_{i}\right|\right)
\label{eq:ham-loss}
\end{equation}
is the mean absolute error between the requested and reconstructed Hamiltonian parameters in min--max-scaled coordinates, and $\mathcal{L}_{\mathrm{range}}$ is a penalty, defined in Appendix~\ref{app:training-details}, that discourages predicted geometries outside the scaled $[0,1]$ design region, with weight $\lambda_{\mathrm{range}}$. We enforce this constraint with a soft penalty instead of a sigmoid output because min--max scaling puts training designs exactly at $0$ and $1$, values a sigmoid can never reach and near which its gradients vanish. Evaluating the loss in scaled coordinates ensures that $f_q$ and $\alpha$, which differ by more than an order of magnitude in physical units, contribute comparably. Architectures and training hyperparameters for both networks were selected using a Keras Tuner \cite{omalley2019kerastuner} search. The search space, selected configuration, and optimizer settings are given in Appendix~\ref{app:training-details}. The best validation loss of 0.0011 came from a single hidden layer of 64 neurons, and increasing depth or width did not improve it (Fig.~\ref{fig:param-sweep}), consistent with the small-data behavior discussed in Section~\ref{subsec:small-data}.

\section{Results}\label{sec:results}

We evaluate the inverse model, trained using the frozen forward surrogate, by passing its predicted geometries through an independent conventional forward-simulation workflow and comparing the recovered Hamiltonian parameters with the requested targets, namely qubit frequency $f_q$ and anharmonicity $\alpha$. We also report results for a separate model trained on the same data with the loss evaluated on the extracted capacitance-matrix elements rather than on the Hamiltonian parameters in Appendix~\ref{app:capacitance-level}.

\subsection{Inverse-plus-surrogate Hamiltonian-level performance}
\label{subsec:results-transmon}

We generate 100 inverse-designed cross-claw transmon candidates, with target $(f_q,\alpha)$ pairs taken from the 291-record held-out test set, and validate them through the conventional-EM/scqubits workflow. Of these, 97 produce usable Hamiltonian results. The remaining three cases violate geometric layout rules (metal overlap or loss of the required ground spacing between claw and cross) rather than failing Hamiltonian reconstruction, and are detailed in Appendix~\ref{app:testing-pipeline}. We therefore report a solver-valid geometry rate of 97\%, and the percent-error statistics below are calculated using the 97 valid designs. In future adaptations of our workflow, unphysical designs could be avoided by adding physics-informed layout penalties or hard geometric constraints that discourage metal overlap, dielectric-gap overlap, and loss of required ground spacing.

\begin{figure*}[!t]
\centering
\captionsetup[subfigure]{labelformat=parens,labelsep=none}
\begin{subfigure}[t]{\columnwidth}
\centering
\includegraphics[width=\linewidth]{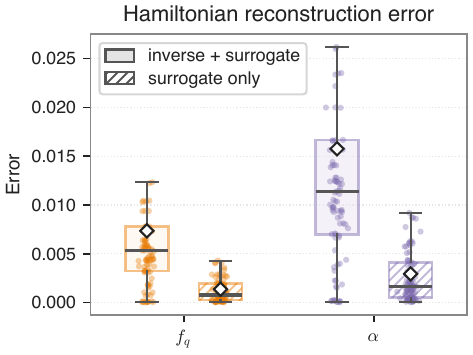}
\caption{}\label{fig:inverse-surrogate-transmon}
\end{subfigure}%
\hfill
\begin{subfigure}[t]{\columnwidth}
\centering
\raisebox{1.2em}{\includegraphics[width=\linewidth]{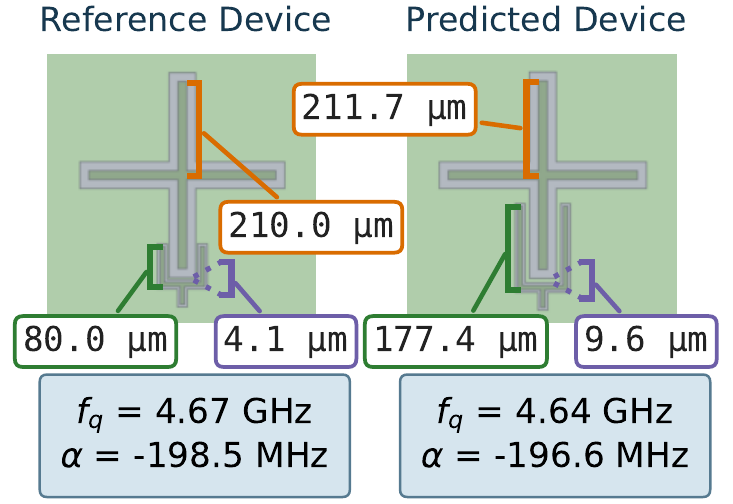}}
\caption{}\label{fig:end2end-example}
\end{subfigure}

\caption{(a) Hamiltonian-level error, reported as a fraction of the reference value, of the combined inverse-plus-surrogate pipeline on the cross-claw transmon (shaded boxes) and of the frozen forward surrogate alone (hatched boxes), for qubit frequency $f_q$ and anharmonicity $\alpha$. Results for the inverse-plus-surrogate and surrogate-only pipeline are shown across the 97 EM-validated designs. Boxes span the interquartile range (IQR), defined as the range from the 25th to the 75th percentile. The heavy line marks the median, the diamond marks the mean, and whiskers extend to the most extreme non-outlier points within $1.5\times$ the IQR. Non-outlier samples overlaid as jittered points. Median (mean) errors for the full pipeline are 0.53\% (0.73\%) for $f_q$ and 1.14\% (1.58\%) for $\alpha$, versus 0.077\% (0.136\%) and 0.165\% (0.293\%) for the surrogate alone. (b) Reference and model-predicted cross-claw transmon layouts for a representative sample. The predicted geometry differs noticeably from the SQuADDS reference, while the reconstructed Hamiltonian closely matches the target, demonstrating the one-to-many nature of the inverse mapping.}
\label{fig:inverse-results-and-example}
\end{figure*}

Across the 97 usable simulation-validated designs, the inverse-plus-surrogate pipeline achieves mean percent errors of 0.73\% for $f_q$ and 1.58\% for $\alpha$, with median error values of 0.53\% and 1.14\% respectively. We report per-sample percent errors because $f_q$ (a few GHz) and $\alpha$ (a few hundred MHz in magnitude) differ in scale by more than an order of magnitude, and we summarize them with box plots (Fig.~\ref{fig:inverse-surrogate-transmon}) because the error distributions are non-Gaussian. Fig.~\ref{fig:inverse-surrogate-transmon} also shows the error of the frozen forward surrogate alone on 97 held-out test samples, with median errors of 0.077\% for $f_q$ and 0.165\% for $\alpha$, indicating that it accurately reproduces EM simulation results on SQuADDS records.
For a 4 to 6\,GHz qubit, this corresponds to frequency errors of a few tens of MHz, while the anharmonicity error is on the order of a few MHz for a typical 150 to 200\,MHz target. These errors reproduce an expected factor-of-two asymmetry. That is, anharmonicity inherits the full fractional capacitance uncertainty directly, while qubit frequency inherits roughly half, since the dominant $\sqrt{8 E_J E_C}$ contribution to $f_q$ suppresses the fractional sensitivity to $E_C$ by a factor of two. Their absolute magnitudes are also within the combined uncertainty of the simulation and fabrication stack as discussed later in Section~\ref{sec:discussion}. In the regions within and adjacent to the SQuADDS dataset evaluated here, further reductions in test-set model error would likely be difficult to observe experimentally unless field-wide reduction in fabrication variability and simulation-to-measurement uncertainty were realized.

\subsection{Runtime and speedup results}\label{subsec:runtime}

\begin{figure*}[!t]
\centering
\captionsetup[subfigure]{labelformat=parens,labelsep=none}
\begin{subfigure}[t]{\columnwidth}
\centering
\includegraphics[width=\columnwidth]{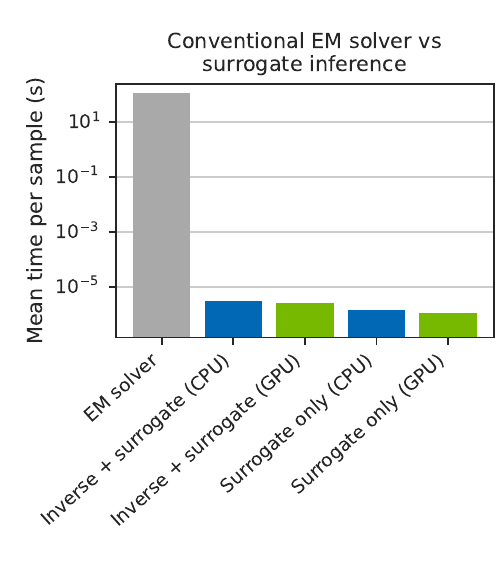}
\caption{}\label{fig:runtime-fastest}
\end{subfigure}
\vspace{0.5em}
\begin{subfigure}[t]{\columnwidth}
\centering
\includegraphics[width=\columnwidth]{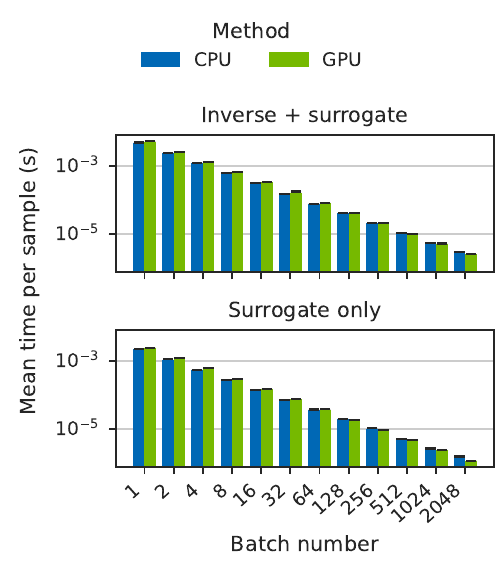}
\caption{}\label{fig:runtime-batch}
\end{subfigure}
\caption{Runtime of the neural pipelines compared with conventional electromagnetic extraction. (a) Per-sample runtime for the conventional EM solver compared with the full inverse-plus-surrogate pipeline and the forward surrogate alone. The neural-pipeline values are reported at batch size 2048, which gives the lowest per-sample runtime among the tested batch sizes. The conventional EM-solver value is the runtime for a single design. Single-design CPU timings for the neural pipelines, which are dominated by the per-call dispatch overhead, are given in Table~\ref{tab:runtime}. The logarithmic time axis highlights the orders-of-magnitude speedup obtained by replacing electromagnetic extraction with neural inference. (b) Batch-size dependence of the mean per-sample neural inference time (averaged over repeated runs at each batch size) for the inverse-plus-surrogate pipeline and the forward surrogate alone, for batch sizes from 1 to 2048. Per-sample cost falls monotonically as the batch grows and TensorFlow dispatch overhead is spread across more samples, so the largest batches are the most efficient. Larger batches still would lower the per-sample cost further, up to memory limits.}
\label{fig:runtime}
\end{figure*}

As electromagnetic simulation is the computational bottleneck in superconducting qubit design loops \cite{bib10}, simulation and design acceleration is a primary motivation of our AI-based approach. To quantify the speedup obtained from the surrogate and inverse-plus-surrogate models, we benchmark the conventional pipeline and the fully neural pipelines on the same cluster, using 8 cores on a single node. The conventional EM solver supports only CPU-based solving, so the CPU timings provide a like-for-like comparison. Neural inference was additionally benchmarked using one NVIDIA A100 SXM4 GPU on the same node for the batch-size sweep in Fig.~\ref{fig:runtime-batch}.

Table~\ref{tab:runtime} separates the conventional electromagnetic-extraction cost from the learned neural-inference cost. Fig.~\ref{fig:runtime-fastest} shows the resulting orders-of-magnitude difference on a logarithmic time scale, and Fig.~\ref{fig:runtime-batch} shows how the neural inference cost decreases with batching.

\begin{table*}[!t]
\caption{Per-design runtime comparison for the conventional EM-solver workflow and the neural inverse-design workflow. Timings were measured on one Quest GPU node using 8 CPU cores, with Python 3.11.15 and TensorFlow 2.20.0, averaged over
50 calls after 10 warm-up calls. Batched values are reported per design. The batch-291 rows use the full held-out test set and are timed through the Keras \texttt{predict()} API, which carries a fixed per-call overhead of roughly 60\,ms independent of batch size. The batch-2048 rows time a direct model call, which removes that overhead. All rows are CPU-only except where the Notes column says GPU. Fig.~\ref{fig:runtime-batch} gives the full batch-size sweep on both
devices.}
\label{tab:runtime}
\centering
\footnotesize
\setlength{\tabcolsep}{3pt}
\renewcommand{\arraystretch}{1.18}
\begin{tabular*}{\textwidth}{@{\extracolsep{\fill}}
p{0.11\textwidth}
p{0.26\textwidth}
p{0.11\textwidth}
p{0.12\textwidth}
p{0.11\textwidth}
p{0.19\textwidth}
@{}}
\toprule
\textbf{Workflow} & \textbf{Stage} & \textbf{Single design} & \textbf{Batched design} & \textbf{Batch size} & \textbf{Notes} \\
\midrule
Conventional & EM capacitance extraction & $\sim$120\,s & $\sim$120\,s & --- & SQuADDS solver runs \\
Conventional & Capacitance-to-Hamiltonian & $<1$\,ms & $<1$\,ms & --- & Analytic/scqubits \\
\midrule
Neural & Inverse MLP & 59\,ms & 0.23\,ms & 291 & Test set \\
Neural & Inverse + surrogate & 60\,ms & 0.24\,ms & 291 & Test set, \texttt{predict()} \\
Neural & Inverse + surrogate & --- & 3.1\,\microsec & 2048 & CPU, direct call \\
Neural & Inverse + surrogate & --- & 2.6\,\microsec & 2048 & GPU, direct call \\ 
\bottomrule
\end{tabular*}
\end{table*}

In particular, a single conventional EM capacitance extraction on the transmon cross takes approximately 2 minutes, which includes design rendering and meshing, actual solver time, and saving results, depending on the geometry and the final mesh size, using SQuADDS simulation parameters with 8 cores. Conversely, a single call to the combined inverse-plus-surrogate pipeline takes about 60\,ms, meaning that generating and screening one candidate design through our learned forward surrogate is approximately three orders of magnitude faster than completing the same task with the conventional EM solver. Furthermore, the single-call number is dominated by TensorFlow per-call dispatch overhead rather than by the actual network math. When the same pipeline is run on batches of samples (as would be the case for virtually any model training protocol), this fixed overhead is spread across the batch and the arithmetic itself runs more efficiently as vectorized batched operations, so the per-sample cost drops to about 0.24\,ms through the Keras \texttt{predict()} API. Timing the model call directly, without that per-call overhead, the same pipeline reaches 3.1\,\microsec per sample on CPU and 2.6\,\microsec per sample on GPU at a batch size of 2048 (Fig.~\ref{fig:runtime-batch}). The forward surrogate alone drops further, to about 1.5\,\microsec per sample on CPU at the same batch size.

This speedup further compounds when the design loop is iterative. For example, the nearest-neighbor stress test in Section~\ref{subsec:small-data} evaluates 50{,}000 candidate designs through the learned surrogate before selecting 100 distance-stratified candidates for EM validation. This entire in-depth search takes only seconds using our surrogate model. In contrast, evaluating all 50{,}000 candidates directly through the conventional EM solver would take over 1{,}500\,hours.

\subsection{Small-data scaling and stress test}\label{subsec:small-data}

Before probing where the surrogate fails, we first investigate how much its accuracy depends on the amount of training data used. We retrain the forward surrogate MLP on uniformly sampled subsets of the training set, from 10\% to 100\% of the available samples, and evaluate each model on the same test set. In order to avoid biasing the smaller fractions with a model architecture selected for the full dataset, the hyperparameters are re-tuned independently for each fraction with a Bayesian Keras Tuner search, and the model is then retrained from scratch with five random seeds.

\begin{figure}[!tbp]
\centering
\includegraphics[width=\columnwidth]{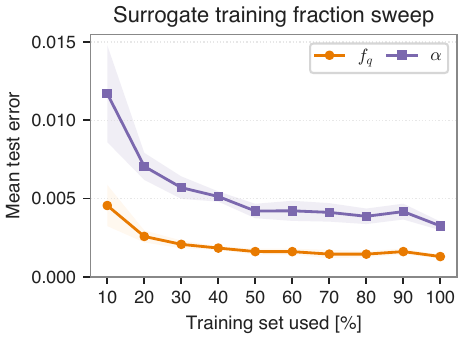}
\caption{Errors for the forward surrogate MLP for the cross-claw transmon at various training data amounts. The surrogate is retrained on uniformly sampled subsets of the training set (10--100\%), with hyperparameters re-tuned independently for each fraction using a Bayesian Keras Tuner search and the selected configuration refit with five random seeds. Lines show the mean held-out test error in qubit frequency $f_q$ (orange) and anharmonicity $\alpha$ (purple), whereas the shaded bands indicate $\pm1\sigma$ across seeds. The error decreases steeply up to roughly half of the training data and then plateaus, indicating that our methodology is highly data-efficient.}
\label{fig:surrogate-data-sweep}
\end{figure}

Fig.~\ref{fig:surrogate-data-sweep} shows the resulting curve. The mean test error drops steeply up to roughly half of the training data, with the $f_q$ error improving from 0.46\% to 0.16\% and the $\alpha$ error from 1.17\% to 0.42\%, and then it largely plateaus, where doubling the data from 50\% to 100\% only brings $f_q$ to 0.13\% and $\alpha$ to 0.33\%. The $\alpha$ error remains roughly twice the $f_q$ error across the sweep, consistent with the capacitance-sensitivity asymmetry noted in Section~\ref{subsec:results-transmon}. Because we reoptimize the architecture at every fraction, the plateau is not the result of fixed hyperparameters. Uniformly adding more samples from the same SQuADDS distribution therefore yields diminishing returns, suggesting that the remaining error is due to where the training data is in design space, rather than by how much data there is.

\begin{figure*}[!t]                                              \centering  
\captionsetup[subfigure]{labelformat=parens,labelsep=none}
\begin{subfigure}[t]{\columnwidth}
\centering
\includegraphics[width=\columnwidth]{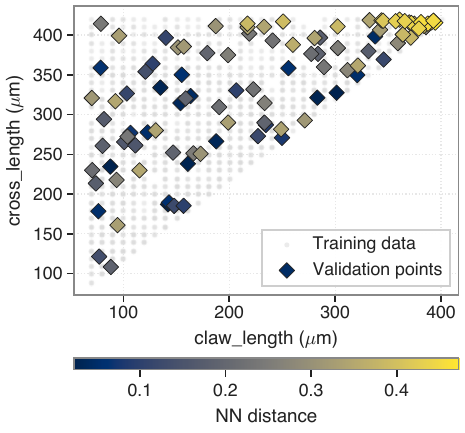}
\caption{}\label{fig:stress-seeds}
\end{subfigure}
\vspace{0.6em}
\begin{subfigure}[t]{\columnwidth}
\centering
\raisebox{1.2em}{\includegraphics[width=\columnwidth]{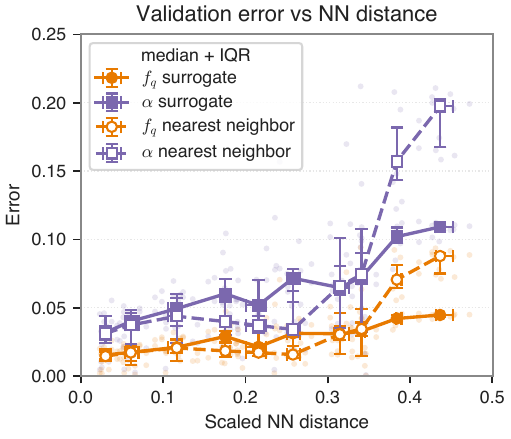}}
\caption{}\label{fig:stress-results}
\end{subfigure}

\caption{Nearest-neighbor stress test of the forward surrogate model. (a) Validation candidates in the claw-length versus cross-length plane of Quantum Metal parameter space are shown to visualize the NN distance test in (b). Grey points show the SQuADDS training distribution, and colored diamonds show the 100 validation candidates, drawn 10 from each of 10 equal-width bins in scaled nearest-training-geometry distance and colored by that distance. All candidates lie within the space spanned by the training data. The underlying pool of 50{,}000 candidates is sampled uniformly, but the plotted subset is deliberately stratified by scaled distance. Large-distance candidates are rare under uniform sampling, so sampling every bin equally oversamples them and necessarily concentrates them in the most sparsely covered region of the training distribution. Larger distances correspond to more sparsely sampled regions, where surrogate reliability is expected to degrade first. The remaining parameter-pair visualizations are shown in Fig.~\ref{fig:stress-pairs-appendix}. (b) Surrogate vs conventional EM solver error as a function of scaled nearest-training-geometry distance, reported as fractions of the EM-solver value, for $f_q$ (orange) and $\alpha$ (purple). Solid lines with filled markers show the surrogate, and dashed lines with open markers show a nearest-neighbor baseline that assigns each candidate the Hamiltonian of its closest training geometry. Markers are placed at the median nearest-training-geometry distance of the 10 candidates in each bin. Vertical error bars give the IQR of the error within a bin, and horizontal bars the IQR of the candidate distances within that bin. Individual candidates are shown as faint points at their own distance, and since each bin spans a wider distance range than its IQR, roughly half of them fall outside the horizontal bars by construction. The surrogate degrades more slowly with distance than the baseline, indicating a smoother in-domain approximation than nearest-neighbor lookup.}
\label{fig:stress-test}
\end{figure*}

To probe that coverage dependence directly, we run a nearest-neighbor stress test. We generate 50{,}000 random cross-claw geometries inside the interpolation region of the training data. Each geometry parameter is first min--max scaled to $[0,1]$ using the training-set minimum and maximum, and $d_{\mathrm{NN}}$ is then the Euclidean distance in that scaled space from a candidate to its closest training geometry, so that small $d_{\mathrm{NN}}$ marks a well-covered region and large $d_{\mathrm{NN}}$ a sparsely supported one. We stratify these candidates into ten equal-width distance bins, draw ten from each bin, and validate the resulting 100 designs against the conventional EM solver. The sampling, distance, and binning definitions are given in Appendix~\ref{app:training-details}, and Fig.~\ref{fig:stress-methodology} summarizes the procedure. Fig.~\ref{fig:stress-seeds} shows where the selected candidates fall in the claw-length versus cross-length plane, colored by their scaled nearest-neighbor distance. The two remaining parameter-pair projections (claw length versus ground spacing, and ground spacing versus cross length) are shown in Fig.~\ref{fig:stress-pairs-appendix} (Appendix~\ref{app:training-details}).

Fig.~\ref{fig:stress-results} compares the surrogate error against the conventional EM solver as a function of $d_{\mathrm{NN}}$, with solid lines and filled markers showing the surrogate and dashed lines with open markers showing a nearest-neighbor baseline that assigns each candidate the Hamiltonian of its closest training geometry. As expected, error grows with distance from the training data, and the anharmonicity error grows about twice as fast as the frequency error, since $|\alpha| \propto E_C \propto 1/C_\Sigma$ inherits the full fractional capacitance error, while the dominant $\sqrt{8E_JE_C}$ contribution to $f_q$ scales as $C_\Sigma^{-1/2}$ and inherits roughly half of it. The two curves agree at small distance, but the surrogate pulls ahead as distance grows. It holds the $f_q$ error to a few percent across the entire tested range, with the $\alpha$ error rising only in regions far from any training support.

If the surrogate were memorizing or locally interpolating between training samples, its error would track the geometric distance as closely as the nearest-neighbor baseline does. Instead it degrades more slowly, indicating that the MLP learns a smoother empirical approximation of the geometry-to-Hamiltonian map than nearest-neighbor lookup provides. This is what makes the pipeline data-efficient, because it interpolates across regions within the training-data domain using a relatively small number of samples, so that even sparsely covered regions can be screened before committing to a full EM validation.

\subsection{End-to-end examples}\label{subsec:end-to-end}

We also check individual designs by passing a single pair of requested Hamiltonian values through the inverse-plus-surrogate pipeline, assembling the resulting cross-claw transmon layout, and validating it with the conventional forward-simulation workflow. Because the inverse map is one-to-many, distinct Quantum Metal geometries can yield nearly identical values of $f_q$ and $\alpha$. The errors we report for this example are inverse-plus-surrogate errors, meaning that they include both the inverse model's geometry prediction and the forward surrogate's reconstruction of the Hamiltonian. Appendix~\ref{app:testing-pipeline} discusses direct solver-in-the-loop validation separately.

For the sample in Fig.~\ref{fig:end2end-example}, the target qubit frequency is 4.67\,GHz and the target anharmonicity is $-198.5$\,MHz. The predicted geometry lands at 4.64\,GHz and $-196.6$\,MHz, giving a 0.64\% error in $f_q$ and a 0.97\% error in $\alpha$, even though the claw length moves from 80.0\,\um to 177.4\,\um and the ground spacing moves from 4.1\,\um to 9.6\,\um, while the cross length increases from 210\,\um to 211.7\,\um. The model appears to be balancing increased cross length and ground spacing (which both increase total capacitance) against increased claw length (which decreases total capacitance by increasing the amount of ground plane removed). The predicted layout should therefore be read as an alternative valid design, not as an attempt to reproduce the SQuADDS reference geometry.

\FloatBarrier
\section{Discussion}\label{sec:discussion}
This work demonstrates a Hamiltonian-targeted inverse-design workflow in which an inverse MLP, trained against a frozen learned forward surrogate, maps target Hamiltonian parameters directly to component-level layout parameters. The ingredients are generic, as the construction requires only a parameterized layout and a simulated dataset linking layout parameters to target quantities, and nothing in the training procedure is specific to Quantum Metal or SQuADDS beyond their role as the layout tool and data source for this first demonstration. On the cross-claw transmon test case, closed-loop validation measures whether the predicted layouts recover the target quantities, with mean percent errors of 0.73\% in $f_q$ and 1.58\% in $\alpha$ across the 97 usable EM-validated designs, corresponding to a 97\% solver-valid geometry rate. We reach this accuracy with on the order of 1{,}000 training samples, so the pipeline is not only accurate but data-efficient, which is invaluable given that each training data point requires an expensive electromagnetic simulation.

We note that such error magnitudes are comparable to the uncertainty of the corresponding simulation and fabrication stack, such that they do not represent a primary or limiting source of device design error \cite{bib10, mohseni2024buildquantumsupercomputerscaling}. Likewise, finite-element simulation carries uncertainties from geometry discretization (meshing), convergence tolerances, and approximate boundary conditions \cite{fong2018finite, Sommers2025SQDMetal}, and these compound with experimental nonidealities such as lithographic variability and junction-area uncertainty \cite{Muthusubramanian2024}. As a result, measured Hamiltonian parameters typically differ from their simulated counterparts at the percent level, with prior work reporting discrepancies in qubit anharmonicity on the order of a few percent \cite{bib7}. Our closed-loop errors are therefore comparable to the smallest differences this benchmark can verify on hardware.

As a result, gains from driving model error further below the simulation-to-measurement gap would be difficult to confirm on individual fabricated devices from a typical academic process, although lower model error can still improve candidate ranking and screening within the design loop. For device-level accuracy, the larger opportunity lies in better targeted coverage of sparse regions of the design space rather than in reducing model error on the existing data alone.

The fully neural pipeline runs in about 60\,ms for a single query on CPU, and in batch mode reconstructs Hamiltonians at about 3\,\microsec per sample on the same CPU class used for the conventional EM benchmark, while a single conventional EM capacitance extraction still requires roughly ${\sim}2$ minutes per sample (Section~\ref{subsec:runtime}). Even so, the more relevant comparison is the full iterative design loop, where reaching a target Hamiltonian typically requires many simulation cycles and substantial designer time. Compared to the kind of iterative design sweep that is routine in SQuADDS-style workflows, the learned pipeline is several orders of magnitude faster per design query, with further gain on GPU. Notably, the model also generates novel, valid layouts, as for many targets it predicts a geometry distinct from the SQuADDS reference that still hits the target Hamiltonian within the simulation-to-measurement uncertainty. This is reflective of the same generative capacity reported across the inverse-design literature \cite{bib3,bib4}. Taken together, the results demonstrate that component-level inverse design is practical for generating and evaluating candidate designs within the sampled design space, even in the small-data regime.

Several directions are natural extensions of this work. Replacing the direct-regression inverse model with a generative model would turn the one-to-many structure of the problem from a training obstacle into a design asset, producing a distribution of valid geometries for each target rather than a single point estimate. Active learning can then direct new EM simulations toward the sparse regions of SQuADDS flagged by our stress test, where each added sample buys the most accuracy, and soft physics-informed penalties can steer the network away from layouts that are infeasible or impractical to fabricate.

Subsequent studies could also target quantum devices beyond the transmon cross. For example, the same Hamiltonian-targeted workflow can be expanded to more diverse and complex superconducting QPU systems. The natural next targets are systems with additional Hamiltonian parameters, such as systems of coupled transmons, coupled cavities, and different qubit modalities, as more diverse validated datasets become available.

\FloatBarrier
\clearpage

\setcounter{mainfigurecount}{\value{figure}}

\begin{appendices}
\setcounter{figure}{\value{mainfigurecount}}
\renewcommand{\thefigure}{\arabic{figure}}

\section{Testing pipeline (EM solver in the loop)}\label{app:testing-pipeline}

\begin{figure*}[!t]
\centering
\includegraphics[width=\textwidth]{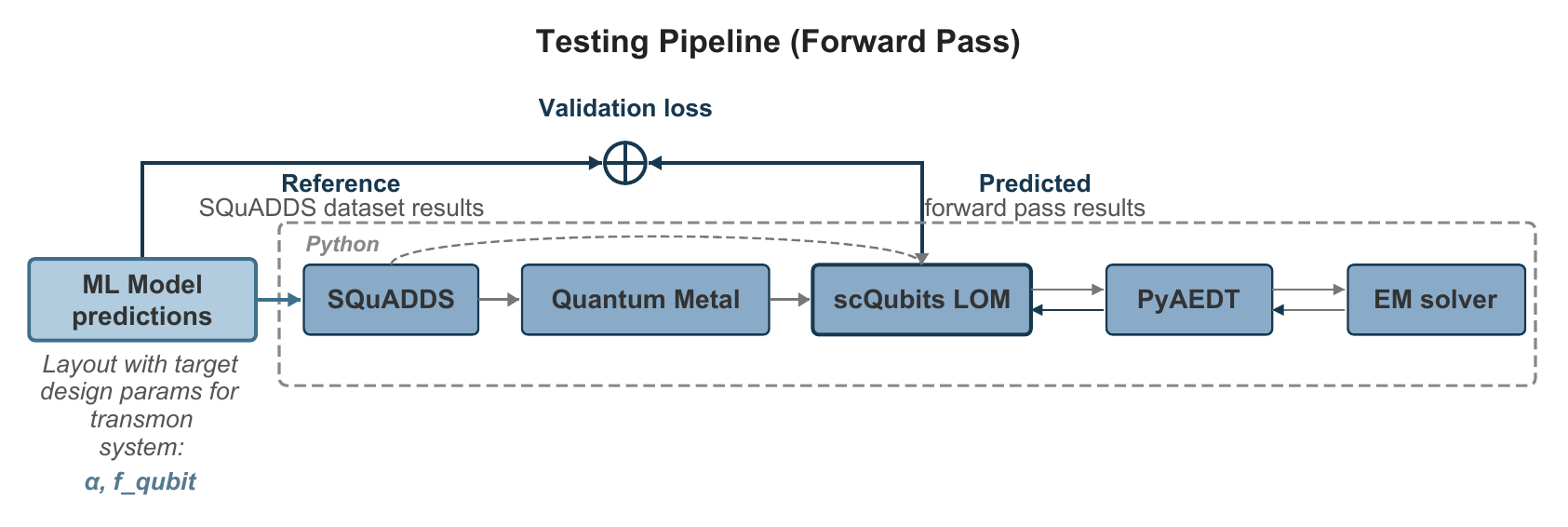}
\caption{Forward-validation pipeline used for forward validation of the inverse model. The model-predicted transmon geometry parameters are passed through the conventional EM-solver workflow to generate forward-pass results. These results are compared with the corresponding SQuADDS reference values to calculate the validation loss. Unlike the training loop, this check keeps the conventional EM solver in the loop and is used only for final validation.}
\label{fig:app-testing-pipeline}
\end{figure*}

Fig.~\ref{fig:app-testing-pipeline} shows the EM-solver-in-the-loop validation path. In this pipeline, the predicted design is passed through the standard Python-to-solver workflow rather than through the learned forward surrogate. The model-predicted Quantum Metal parameters are first rendered into a layout. Then, the capacitance matrix is extracted with the conventional EM solver. The Hamiltonian parameters are then recovered with the scqubits Lumped Oscillator Model (LOM) analysis that was used in building the SQuADDS dataset. These recovered $f_q$ and $\alpha$ values are then compared against the targets that were requested from the model. This path is slower than the surrogate check by roughly three orders of magnitude, as reported in Section~\ref{subsec:runtime}, therefore we only use it for final validation of candidate designs, rather than during training. 

Three of the 100 inverse-designed candidates in Section~\ref{subsec:results-transmon} failed this validation path and are excluded from the reported percent-error statistics. One candidate failed during EM simulation because the predicted claw arm was long enough for the claw center conductor to overlap the cross-claw transmon center conductor, producing overlapping metal regions. Two additional candidates were removed during analysis because the claw dielectric gap overlapped the qubit dielectric gap, leaving no valid ground spacing between the claw and the cross. All three are geometric layout-rule violations rather than Hamiltonian-reconstruction failures.

\section{Inverse-only model (without surrogate)}\label{app:inverse-only-transmon}

For completeness, this Appendix reports how the inverse model performs when trained against the extracted capacitance-matrix elements rather than against the reconstructed Hamiltonian parameters, without the forward surrogate in the training loop. The headline surrogate result in the main text Section~\ref{subsec:results-transmon} (Fig.~\ref{fig:inverse-surrogate-transmon}), is the stronger demonstration, so we retain this capacitance-level model here as a reference point.

\begin{figure}[!tbp]
\centering
\includegraphics[width=\columnwidth]{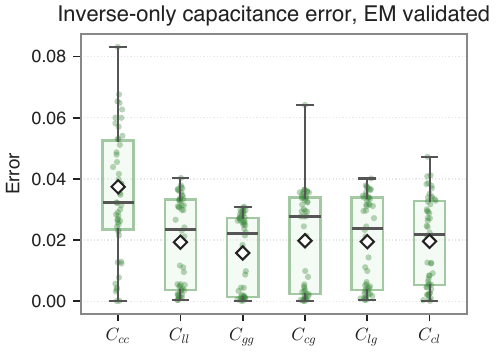}
\caption{Capacitance-matrix errors for the inverse-only model, trained against the capacitance-matrix elements, without the forward surrogate in the training loop. Predicted capacitances are compared against the SQuADDS reference values for the same target. Subscripts denote the cross (c), claw (l), and ground (g) conductors. Boxes span the IQR, with the median marked by the heavy line, the mean by the diamond, whiskers at $1.5\times$ the IQR, and non-outlier designs overlaid as jittered points. Median errors are 2--3\% across all six matrix elements.}
\label{fig:app-inverse-only-transmon-a}
\end{figure}

\begin{figure}[!tbp]
\centering
\includegraphics[width=\columnwidth]{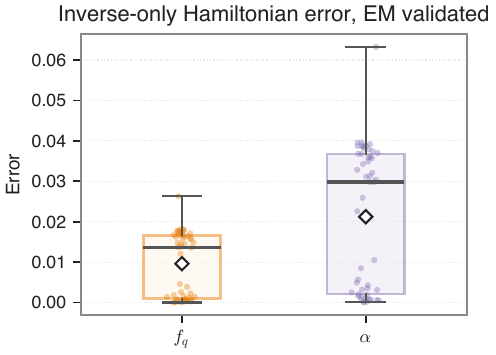}
\caption{Hamiltonian-level errors for the same 50 EM-validated inverse-only designs as in Fig.~\ref{fig:app-inverse-only-transmon-a}, for qubit frequency $f_q$ and anharmonicity $\alpha$, relative to the SQuADDS reference values. Box conventions as in Fig.~\ref{fig:app-inverse-only-transmon-a}. Median errors are 1.4\% for $f_q$ and 3.0\% for $\alpha$.}
\label{fig:app-inverse-only-transmon-b}
\end{figure}

Fig.~\ref{fig:app-inverse-only-transmon-a} shows the error of the predicted capacitance matrix, and Fig.~\ref{fig:app-inverse-only-transmon-b} shows the corresponding qubit frequency and anharmonicity errors, for 50 designs whose targets are drawn from held-out SQuADDS records. Predicted capacitances are compared element by element against the SQuADDS reference values, and the Hamiltonian parameters are recovered from those predicted capacitances through the scqubits LOM conversion of Fig.~\ref{fig:app-testing-pipeline}. The model reproduces the reference capacitances to median errors of roughly 2--3\% across the six matrix entries, which carries through to the Hamiltonian, where $f_q$ and $\alpha$ agree to median errors of 1.4\% and 3.0\%, respectively. Because this model is graded on capacitance values, it has no mechanism to trade off individual matrix elements against one another in service of the resulting Hamiltonian, which is what the surrogate-defined loss in the main text provides.

\section{Capacitance comparison}\label{app:capacitance-level}

\begin{figure}[!tbp]
\centering
\includegraphics[width=\columnwidth]{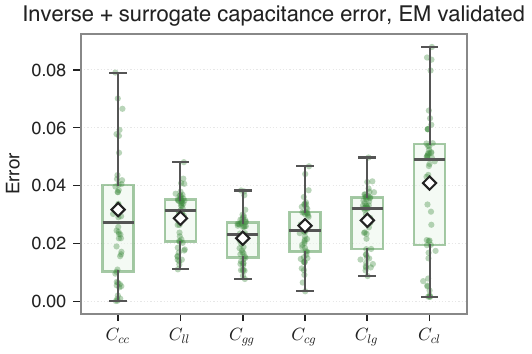}
\caption{Capacitance-level validation of the inverse-plus-surrogate model trained against capacitance-matrix elements, for 50 EM-validated designs. The fractional error of each capacitance-matrix element is shown relative to the SQuADDS reference value for the same target.}
\label{fig:app-capacitance-level}
\end{figure}

Fig.~\ref{fig:app-capacitance-level} shows the distribution of capacitance-matrix errors for the 50 EM-validated designs produced by the inverse-plus-surrogate model trained against capacitance-matrix elements rather than Hamiltonian parameters, the separate model referred to in Section~\ref{sec:results}. The tandem reconstructs the requested capacitances to a median error of 0.31\%, and the layouts it produces realize those capacitances to a median of 2.7\% once rendered and re-simulated with the conventional EM solver. Agreement is comparable across the five larger matrix elements, including the cross-to-ground term that dominates $C_\Sigma$ and therefore sets $E_C$ (2.4\% median). The small cross-to-claw coupling is the least well reproduced at 4.9\%, as expected for the smallest element of the matrix. The gap between the reconstruction error and the realized error reflects that the predicted geometry must additionally survive rendering and re-simulation, so the surrogate is necessarily more accurate than the design it ultimately produces.

\section{Training-data coverage, hyperparameter sweep, and additional stress-test projections}\label{app:training-details}

\begin{figure*}[!t]
\centering
\includegraphics[width=0.95\textwidth]{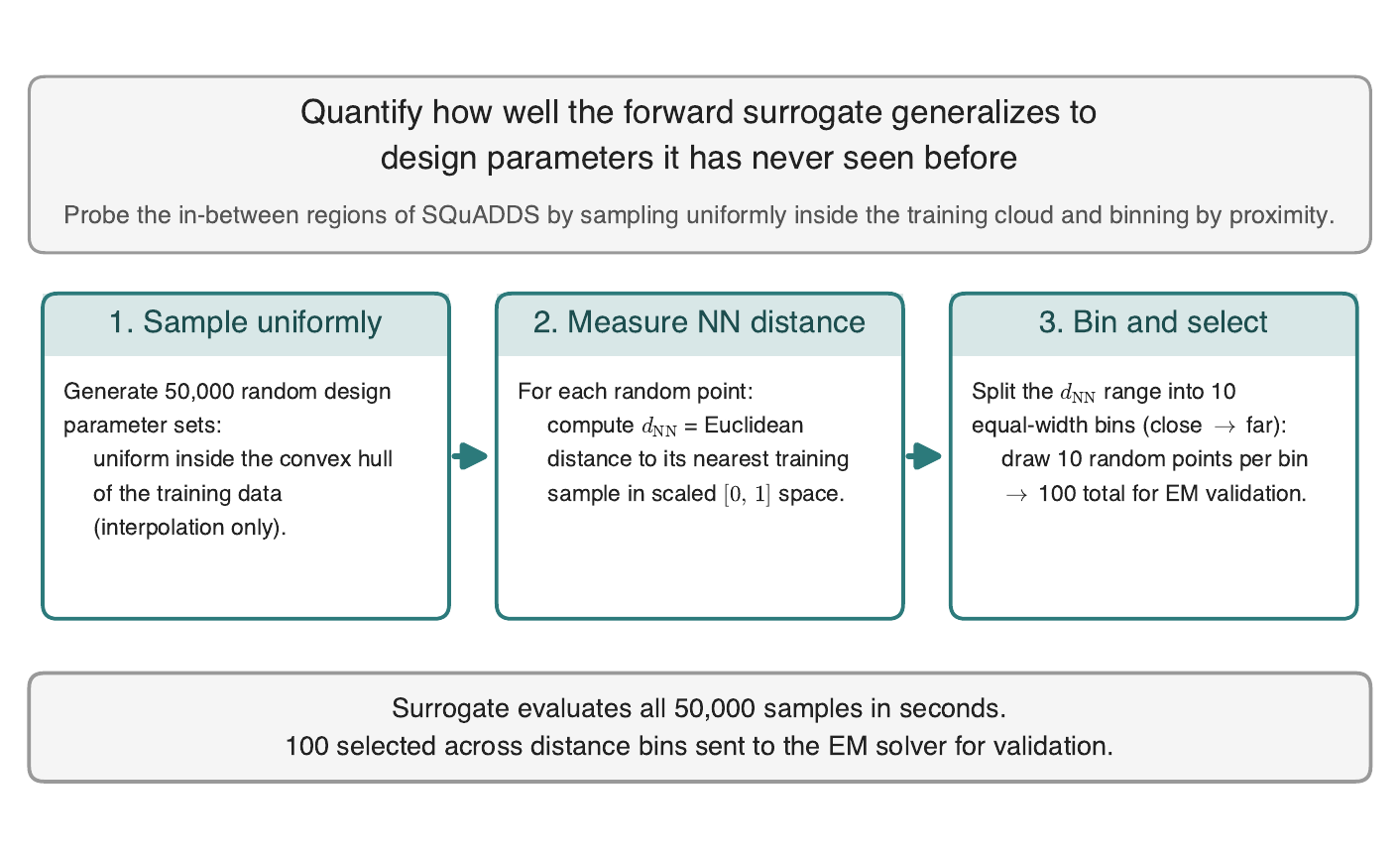}
\caption{Nearest-neighbor stress-test methodology. Uniformly random Quantum Metal parameter sets are generated inside the cross-claw transmon training data space, constrained to remain inside of the interpolated SQuADDS design domain. The learned surrogate evaluates 50{,}000 valid candidates, and each candidate is assigned the Euclidean distance to its nearest training sample, measured after each geometry parameter is min--max scaled to $[0,1]$ on the training set. The candidates are then split into 10 equal-width nearest-neighbor-distance bins, and 10 candidates are selected from each bin for validation, giving 100 validation designs in total. Because training points are scattered throughout this region, most uniform candidates fall close to one, so the far-distance bins are sparsely populated. Drawing equally from every bin therefore deliberately oversamples the far-from-training region, which is what produces the clustering visible in Figs.~\ref{fig:stress-seeds} and~\ref{fig:stress-pairs-appendix}.}
\label{fig:stress-methodology}
\end{figure*}

\begin{figure}[!tbp]
\centering
\includegraphics[width=0.95\columnwidth]{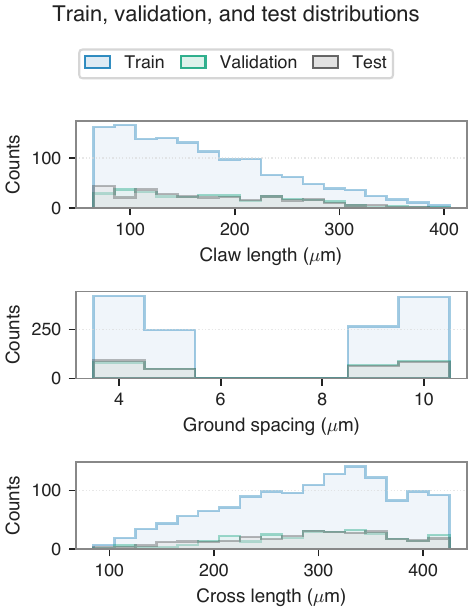}
\caption{Distributions of the three Quantum Metal output parameters used for the inverse model, separated by train, validation, and test split, following the fixed 70/15/15 partition
used throughout training and evaluation.}
\label{fig:sample-data-distribution}
\end{figure}

\begin{figure}[!tbp]
\centering
\includegraphics[width=\columnwidth]{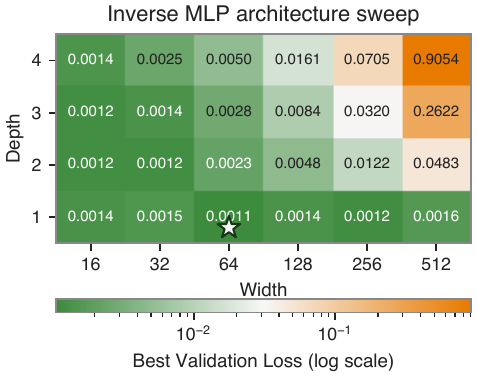}
\caption{Architecture sweep for the inverse MLP over the displayed combinations of depth and width. Each cell shows the best validation loss obtained for that architecture, with the color scale shown logarithmically. The white star marks the lowest loss in this sweep, found for one hidden layer of width 64.}
\label{fig:param-sweep}
\end{figure}

\begin{figure}[!tbp]
\centering
\includegraphics[width=\columnwidth]{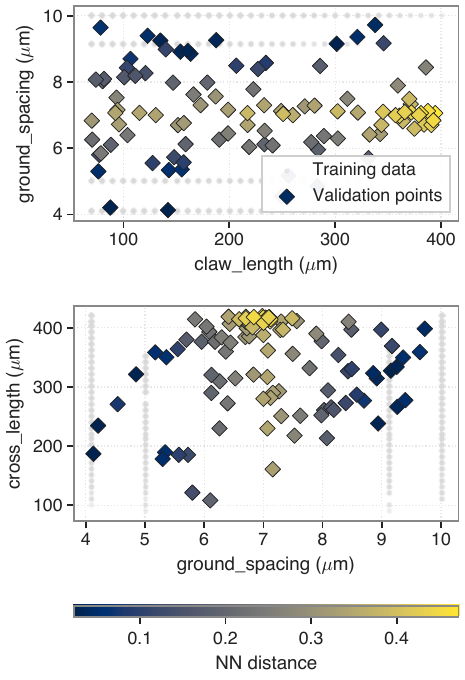}
\caption{Remaining parameter-pair projections of the nearest-neighbor stress test in Quantum Metal parameter space, complementing Fig.~\ref{fig:stress-seeds}. The grey points show the SQuADDS
training distribution, and colored points show the 100 stratified validation candidates (10 per nearest-neighbor-distance bin) colored by their scaled nearest-training-sample distance. The clustering of large-distance candidates reflects the selection process, which draws equally from every distance bin, rather than the uniform sampling of the underlying 50{,}000-candidate dataset.}
\label{fig:stress-pairs-appendix}
\end{figure}

This Appendix collects supporting detail referenced in the main text. This includes the split-wise training-data distributions (Fig.~\ref{fig:sample-data-distribution}), the hyperparameter sweep used to select the inverse-model architecture (Fig.~\ref{fig:param-sweep}), the sampling and candidate-selection procedure behind the nearest-neighbor stress test of Section~\ref{subsec:small-data} (Fig.~\ref{fig:stress-methodology}), and the remaining parameter-pair projections of that test (Fig.~\ref{fig:stress-pairs-appendix}).

For the nearest-neighbor stress test of Section~\ref{subsec:small-data}, candidate geometries are sampled only from the interpolation domain,
\begin{equation}
\mathbf{y}^{(r)} \in \mathrm{conv}\left(\{\mathbf{y}^{(n)}_{\mathrm{train}}\}_{n=1}^{N_{\mathrm{train}}}\right)
\label{eq:convex-hull-condition}
\end{equation}
and are additionally required to satisfy the cross-claw transmon non-overlap constraint
\begin{equation}
\ell_{\mathrm{claw}} + 2g_{\mathrm{claw}} < s_{\mathrm{ground}} + \ell_{\mathrm{cross}}
\label{eq:geometry-validity}
\end{equation}
where $g_{\mathrm{claw}}$ is the fixed claw-gap parameter. This constraint rejects layouts in which the coupling claw would overlap the transmon cross. For each valid candidate $\tilde{\mathbf{y}}^{(r)}$ (min--max scaled), we define the nearest-training-sample distance
\begin{equation}
\begin{aligned}
&d_{\mathrm{NN}}(\tilde{\mathbf{y}}^{(r)})\\
&\quad=\min_{1 \le n \le N_{\mathrm{train}}}\sqrt{\sum_{j=1}^{N_y}\left(\tilde{y}^{(r)}_j-\tilde{y}^{(n)}_{\mathrm{train},j}\right)^2}
\end{aligned}
\label{eq:nearest-neighbor-distance}
\end{equation}
with $N_y=3$; computing it in scaled $[0,1]$ coordinates lets claw length, ground spacing, and cross length contribute on comparable numerical scales. We divide the candidate set into $N_{\mathrm{bin}}=10$ equal-width bins with edges
\begin{equation}
\begin{aligned}
b_m &= d_{\min} + m\,\frac{d_{\max}-d_{\min}}{N_{\mathrm{bin}}},\\
&\quad m = 0,\ldots,N_{\mathrm{bin}}
\end{aligned}
\label{eq:nn-bin-edges}
\end{equation}
where $d_{\min}$ and $d_{\max}$ are the extreme nearest-neighbor distances among the valid candidates, assigning candidate $r$ to bin $m$ when $b_{m-1}\le d_{\mathrm{NN}}(\tilde{\mathbf{y}}^{(r)}) < b_m$, and draw 10 candidates per bin for a total of 100 EM-validated designs. For each, we report the fractional surrogate-to-EM percent error
\begin{equation}
\epsilon^{\text{surr-EM}}_{rk} = \frac{\bigl|\hat{h}^{\text{surr}}_{rk} -h^{\text{EM}}_{rk}\bigr|}{\bigl|h^{\text{EM}}_{rk}\bigr|}
\label{eq:surrogate-EM-error}
\end{equation}
where $r$ indexes the candidate and $k$ the Hamiltonian parameter. The absolute value in the denominator keeps the anharmonicity error, for which $h^{\text{EM}}_{rk} < 0$, positive and on the same footing as the frequency error. Figure~\ref{fig:stress-results} plots $\epsilon^{\text{surr-EM}}_{rk}$ directly, and percentages quoted in the text are $100\,\epsilon^{\text{surr-EM}}_{rk}$.

For the reconstruction term in $\mathcal{L}_{\mathrm{total}}$, we tested both the mean squared error (MSE) and mean absolute error (MAE) \cite{hastie2009elements,sklearn_regression_metrics}. For targets $h_{ik}$ and predictions $\hat{h}_{ik}$, where $i$ indexes the sample and $k$ indexes the Hamiltonian parameter as in Eq.~\eqref{eq:surrogate-EM-error}, these are $\mathrm{MSE}=\frac{1}{N N_h}\sum_{i=1}^{N}\sum_{k=1}^{N_h}(h_{ik}-\hat{h}_{ik})^2$ and $\mathrm{MAE}=\frac{1}{N N_h}\sum_{i=1}^{N}\sum_{k=1}^{N_h}|h_{ik}-\hat{h}_{ik}|$, with $N_h=2$ for the two Hamiltonian targets. MSE assigns disproportionate weight to the few samples in sparsely covered regions of the SQuADDS parameter space because the residuals are squared, whereas MAE weights residuals linearly. MAE gave slightly more stable training and better performance. The range penalty in $\mathcal{L}_{\mathrm{total}}$ acts on the predicted min--max-scaled geometry outputs,
\begin{equation}
\begin{aligned}
\mathcal{L}_{\mathrm{range}}=\frac{1}{N_y}\sum_{j=1}^{N_y}\Big[
&\max(0,-\hat{y}_j)^2\\
&{}+\max(0,\hat{y}_j-1)^2\Big]
\end{aligned}
\label{eq:range-penalty}
\end{equation}
where $N_y=3$ for the cross-claw transmon geometry vector. The range penalty is distinct from L2 regularization, as it acts on the predicted geometry outputs, whereas L2 regularization acts on the network weights.

We used Keras Tuner \cite{omalley2019kerastuner} to optimize over network depth ($\in \{1,\ldots,5\}$ hidden layers), hidden-layer width ($\in [64, 1024]$ neurons per layer), learning rate ($\in [10^{-4}, 10^{-1}]$), L2 regularization coefficient, dropout rate, range-penalty weight, and the inclusion of batch normalization after hidden dense layers. The trainable parameter count follows from the chosen depth and width and was not tuned independently. The forward surrogate was trained with Adam at an initial learning rate of $9.82\times10^{-4}$, batch size 128, and MAE loss. The inverse model was then trained with the surrogate frozen, using a learning rate of $1.0\times10^{-3}$, L2 coefficient $10^{-6}$, dropout 0, range-penalty weight 1, batch size 128, and at most 400 epochs. Both used early stopping on validation loss with a patience of 60 epochs. As a separate exploratory check, Fig.~\ref{fig:param-sweep} shows the validation loss across a smaller grid of model depths and widths. If capacity were the main bottleneck, larger models would reduce validation loss. Instead, the loss flattens and then worsens with increasing depth or width, consistent with the small-data behavior discussed in Section~\ref{subsec:small-data}.


\clearpage
\end{appendices}

\backmatter

\bmhead{Supplementary information}

No supplementary information is included with this version.

\bmhead{Acknowledgements}
The authors would like to thank Samuel Stein, Ang Li, and Chenxu Liu at Pacific Northwest National Laboratory for early discussions regarding this work. Olivia Seidel would also like to thank Giuseppe Di Guglielmo for valuable mentoring regarding the construction and use of an MLP for a previous project, which was later applied to this work. The authors used Anthropic's Claude (Opus 4.7) only for language editing and clarity checks. All technical content, analysis, and conclusions are the authors' own. This manuscript has been coauthored by FermiForward Discovery Group, LLC under Contract No. 89243024CSC000002 with the U.S. Department of Energy, Office of Science, Office of High Energy Physics.

\section*{Declarations}

\begin{itemize}
\item \textbf{Funding} Funding was provided by the Laboratory for Physical Sciences and the Army Research Office under grant W911NF-25-1-0255. O.S. was supported by a Graduate Instrumentation Research Award (GIRA) through the Coordinating Panel for Advanced Detectors (CPAD) as funded by the DOE Office of High Energy Physics. This research was supported in part through the computational resources and staff contributions provided for the Quest high-performance computing facility at Northwestern University which is jointly supported by the Office of the Provost, the Office for Research, and Northwestern University Information Technology.
\item \textbf{Conflict of interest / Competing interests} The authors declare that they have no competing interests.
\item \textbf{Ethics approval and consent to participate} Not applicable. This work is a computational and simulation study and did not involve human participants, their data, or animal subjects.
\item \textbf{Consent for publication} Not applicable. This manuscript does not contain data from any individual person.
\item \textbf{Data availability} The datasets used in this study are derived from SQuADDS. The processed training, validation, and test splits, fitted scalers, and trained model weights are available in the code repository referenced under Code availability.
\item \textbf{Materials availability} Not applicable. This is a computational study and generated no physical materials; the datasets and code are covered by the Data availability and Code availability statements above.
\item \textbf{Code availability} The code used to produce the results in this work is available at \url{https://github.com/CosmiQuantum/ML_qubit_design}.
\item \textbf{Author contribution} 
ELF, SAS, and SS conceptualized using ML for qubit design, using the SQuADDS database as training data, and the Hamiltonian-targeted inverse-design framing. OS conceptualized using an MLP for this task, and performed the model training and tuning using TensorFlow and Keras. TP conceptualized pairing it with a frozen surrogate to address the one-to-many mapping problem, along with the nearest neighbor stress test and defining the loss on the Hamiltonian values. FA performed the validation of the model outputs and stress tests, helped in benchmarking the compute timing tests along with OS, and did model training and hyperparameter tuning as well. SS also provided feedback and assistance with building the validation pipeline. The analysis was done by OS and FA, with feedback from SS, TP, ELF, NP, EFF that adjusted the direction of analysis accordingly. JA and DB provided advising and general project guidance. DB also contributed to reviewing and editing the manuscript. SAS, ELF, and SD were heavily involved in the making of training data that this work is based on. AC helped set up the validation pipeline. BK provided guidance on AI and manuscript. OS wrote the initial draft, and ELF, TP, SS, SAS, HY, and DB helped with the editing of this manuscript. ELF, SS, and EFF helped with the funding acquisition. All authors reviewed and approved the final manuscript.
\end{itemize}

\bibliography{sn-bibliography}

@article{bib1,
author		= "Koch, J. and Yu, T. M. and Gambetta, J. and Houck, A. A. 
                and Schuster, D. I. and Majer, J. and Blais, A. 
                and Devoret, M. H. and Girvin, S. M. and Schoelkopf, R. J.",
title			= "Charge-insensitive qubit design derived from the {C}ooper pair box",
journal		= "Phys. Rev. A",
volume		= "76",
number		= "4",
pages			= "042319",
year			= "2007",
doi			= "10.1103/PhysRevA.76.042319"
}

@article{bib2,
author		= "Peurifoy, J. and Shen, Y. and Jing, L. and Yang, Y. 
                and Cano-Renteria, F. and DeLacy, B. G. 
                and Joannopoulos, J. D. and Tegmark, M. and Soljacic, M.",
title			= "Nanophotonic particle simulation and inverse design using 
                artificial neural networks",
journal		= "Sci. Adv.",
volume		= "4",
number		= "6",
pages			= "eaar4206",
year			= "2018",
doi			= "10.1126/sciadv.aar4206"
}

@article{bib3,
author		= "Liu, D. and Tan, Y. and Khoram, E. and Yu, Z.",
title			= "Training deep neural networks for the inverse design of 
                nanophotonic structures",
journal		= "ACS Photonics",
volume		= "5",
number		= "4",
pages			= "1365--1369",
year			= "2018",
doi			= "10.1021/acsphotonics.7b01377"
}

@article{bib4,
author		= "Molesky, S. and Lin, Z. and Piggott, A. Y. and Jin, W. 
                and Vuckovic, J. and Rodriguez, A. W.",
title			= "Inverse design in nanophotonics",
journal		= "Nat. Photonics",
volume		= "12",
number		= "11",
pages			= "659--670",
year			= "2018",
doi			= "10.1038/s41566-018-0246-9"
}

@article{bib5,
author  = {Wong, Hiu Yung and Dhillon, Prabjot and Beck, Kristin M. and Rosen, Yaniv J.},
title   = {A simulation methodology for superconducting qubit readout fidelity},
journal = {Solid-State Electronics},
volume  = {201},
pages   = {108582},
year    = {2023},
doi     = {10.1016/j.sse.2022.108582}
}

@inproceedings{bib6,
author    = {Lu, Albert and Wong, Hiu Yung},
title     = {Rapid simulation framework for superconducting qubit readout system inverse design and optimization},
booktitle = {2024 International Conference on Simulation of Semiconductor Processes and Devices (SISPAD)},
pages     = {1--4},
year      = {2024},
doi       = {10.1109/SISPAD62626.2024.10733335}
}

@article{bib7,
author		= "Shanto, S. and Kuo, A. and Miyamoto, C. and Zhang, H. 
                and Maurya, V. and Vlachos, E. and Hecht, M. and Shum, C. W. 
                and Levenson-Falk, E. M.",
title			= "{SQ}u{ADDS}: a validated design database and simulation workflow 
                for superconducting qubit design",
journal		= "Quantum",
volume		= "8",
pages			= "1465",
year			= "2024",
doi			= "10.22331/q-2024-09-09-1465"
}

@misc{bib9,
title        = {Qiskit Metal},
year         = {2026},
howpublished = {\url{https://qiskit-community.github.io/qiskit-metal/}},
note         = {Accessed 2026-07-21}
}

@article{bib10,
title         = {A review of design concerns in superconducting quantum circuits},
author        = {Levenson-Falk, Eli M. and Shanto, Sadman Ahmed},
journal       = {Materials for Quantum Technology},
volume        = {5},
number        = {2},
pages         = {022003},
year          = {2025},
doi           = {10.1088/2633-4356/ade10d},
eprint        = {2411.16967},
archivePrefix = {arXiv},
primaryClass  = {quant-ph}
}

@article{bib12,
title = {Energy-participation quantization of Josephson circuits},
author = {Minev, Zlatko K. and Leghtas, Zaki and Mundhada, Shantanu O. and Christakis, Lysander and Pop, Ioan M. and Devoret, Michel H.},
journal = {npj Quantum Information},
volume = {7},
number = {1},
pages = {131},
year = {2021},
doi = {10.1038/s41534-021-00461-8}
}

@article{bib13,
title={scqubits: a Python package for superconducting qubits},
author={Groszkowski, Peter and Koch, Jens},
journal={Quantum},
volume={5},
pages={583},
year={2021},
publisher={Verein zur F{\"o}rderung des Open Access Publizierens in den Quantenwissenschaften},
doi={10.22331/q-2021-11-17-583}
}

@article{Muthusubramanian2024,
author  = {Muthusubramanian, N. and Finkel, M. and Duivestein, P. and Zachariadis, C. and van der Meer, S. L. M. and Veen, H. M. and Beekman, M. W. and Stavenga, T. and Bruno, A. and DiCarlo, L.},
title   = {Wafer-scale uniformity of {D}olan-bridge and bridgeless {M}anhattan-style {J}osephson junctions for superconducting quantum processors},
journal = {Quantum Science and Technology},
volume  = {9},
number  = {2},
pages   = {025006},
year    = {2024},
doi     = {10.1088/2058-9565/ad199c}
}

@article{yang2025deep,
title={Deep Generative Prior for First Order Inverse Optimization},
author={Yang, Haoyu and Azizzadenesheli, Kamyar and Ren, Haoxing},
journal={arXiv preprint arXiv:2504.20278},
year={2025}
}

@article{mcmc,
title={MCMC methods for functions: modifying old algorithms to make them faster},
author={Cotter, Simon L and Roberts, Gareth O and Stuart, Andrew M and White, David},
journal={Statistical Science},
volume={28},
number={3},
pages={424--446},
year={2013},
publisher={JSTOR}
}

@article{fno,
title={Fourier neural operator for parametric partial differential equations},
author={Li, Zongyi and Kovachki, Nikola and Azizzadenesheli, Kamyar and Liu, Burigede and Bhattacharya, Kaushik and Stuart, Andrew and Anandkumar, Anima},
journal={arXiv preprint arXiv:2010.08895},
year={2020}
}

@article{huang2024data,
title={Data-and Physics-driven Deep Learning Based Reconstruction for Fast MRI: Fundamentals and Methodologies},
author={Huang, Jiahao and Wu, Yinzhe and Wang, Fanwen and Fang, Yingying and Nan, Yang and Alkan, Cagan and Abraham, Daniel and Liao, Congyu and Xu, Lei and Gao, Zhifan and others},
journal={IEEE Reviews in Biomedical Engineering},
year={2024}
}

@article{long2024invertible,
title={Invertible Fourier Neural Operators for Tackling Both Forward and Inverse Problems},
author={Long, Da and Xu, Zhitong and Yuan, Qiwei and Yang, Yin and Zhe, Shandian},
journal={arXiv preprint arXiv:2402.11722},
year={2024}
}

@inproceedings{NugrahaShao2023,
author    = {Nugraha, Ferris Prima and Shao, Qiming},
title     = {Machine Learning-Based Predictive Modeling for Designing Transmon Superconducting Qubits},
booktitle = {2023 IEEE International Conference on Quantum Computing and Engineering (QCE)},
pages     = {1360--1368},
year      = {2023},
publisher = {IEEE},
doi       = {10.1109/QCE57702.2023.00154},
address = {Bellevue, WA, USA}
}

@article{Blais2004CircuitQED,
author  = {Blais, Alexandre and Huang, Ren-Shou and Wallraff, Andreas and Girvin, S. M. and Schoelkopf, R. J.},
title   = {Cavity quantum electrodynamics for superconducting electrical circuits: An architecture for quantum computation},
journal = {Physical Review A},
volume  = {69},
number  = {6},
pages   = {062320},
year    = {2004},
doi     = {10.1103/PhysRevA.69.062320}
}

@article{Devoret2013SuperconductingCircuits,
author  = {Devoret, M. H. and Schoelkopf, R. J.},
title   = {Superconducting Circuits for Quantum Information: An Outlook},
journal = {Science},
volume  = {339},
number  = {6124},
pages   = {1169--1174},
year    = {2013},
doi     = {10.1126/science.1231930}
}

@article{Krantz2019QuantumEngineer,
author  = {Krantz, Philip and Kjaergaard, Morten and Yan, Fei and Orlando, Terry P. and Gustavsson, Simon and Oliver, William D.},
title   = {A quantum engineer's guide to superconducting qubits},
journal = {Applied Physics Reviews},
volume  = {6},
number  = {2},
pages   = {021318},
year    = {2019},
doi     = {10.1063/1.5089550}
}

@article{Kjaergaard2020SuperconductingQubits,
author  = {Kjaergaard, Morten and Schwartz, Mollie E. and Braum{\"u}ller, Jochen and Krantz, Philip and Wang, Joel I.-J. and Gustavsson, Simon and Oliver, William D.},
title   = {Superconducting Qubits: Current State of Play},
journal = {Annual Review of Condensed Matter Physics},
volume  = {11},
pages   = {369--395},
year    = {2020},
doi     = {10.1146/annurev-conmatphys-031119-050605}
}

@article{Blais2021CircuitQED,
author  = {Blais, Alexandre and Grimsmo, Arne L. and Girvin, S. M. and Wallraff, Andreas},
title   = {Circuit quantum electrodynamics},
journal = {Reviews of Modern Physics},
volume  = {93},
number  = {2},
pages   = {025005},
year    = {2021},
doi     = {10.1103/RevModPhys.93.025005}
}

@article{Nigg2012BlackBox,
author  = {Nigg, Simon E. and Paik, Hanhee and Vlastakis, Brian and Kirchmair, Gerhard and Shankar, Shyam and Frunzio, Luigi and Devoret, Michel H. and Schoelkopf, Robert J. and Girvin, S. M.},
title   = {Black-Box Superconducting Circuit Quantization},
journal = {Physical Review Letters},
volume  = {108},
number  = {24},
pages   = {240502},
year    = {2012},
doi     = {10.1103/PhysRevLett.108.240502}
}

@misc{Wang2021QiskitMetal,
author       = {Minev, Zlatko K. and McConkey, Thomas G. and Drysdale, J. and Shah, P. and Wang, D. and Facchini, M. and Harper, G. and Blair, J. and Zhang, H. and Lanzillo, N. and Mukesh, S. and Shanks, W. and Warren, C. and Gambetta, Jay M.},
title        = {Qiskit Metal: An Open-Source Framework for Quantum Device Design and Analysis},
year         = {2021},
publisher    = {Zenodo},
doi          = {10.5281/zenodo.4618154}
}

@article{Menke2021Automated,
author  = {Menke, Tim and H{\"a}se, Florian and Gustavsson, Simon and Kerman, Andrew J. and Oliver, William D. and Aspuru-Guzik, Al{\'a}n},
title   = {Automated design of superconducting circuits and its application to 4-local couplers},
journal = {npj Quantum Information},
year    = {2021},
volume  = {7},
number  = {1},
pages   = {49},
doi     = {10.1038/s41534-021-00382-6},
url     = {https://doi.org/10.1038/s41534-021-00382-6}
}

@misc{Sommers2025SQDMetal,
title         = {Open-Source Highly Parallel Electromagnetic Simulations for Superconducting Circuits},
author        = {Sommers, David and Degnan, Zach and Gautam, Divita and Chen, Yi-Hsun and Chiu, Chun-Ching and Fedorov, Arkady and Pakkiam, Prasanna},
year          = {2025},
eprint        = {2511.01220},
archivePrefix = {arXiv},
primaryClass  = {quant-ph},
url           = {https://arxiv.org/abs/2511.01220}
}

@book{hastie2009elements,
title     = {The Elements of Statistical Learning: Data Mining, Inference, and Prediction},
author    = {Hastie, Trevor and Tibshirani, Robert and Friedman, Jerome},
edition   = {2},
year      = {2009},
publisher = {Springer},
address   = {New York}
}

@misc{sklearn_regression_metrics,
title        = {Scikit-learn User Guide: Regression metrics},
author       = {{scikit-learn developers}},
year         = {2026},
howpublished = {\url{https://scikit-learn.org/stable/modules/model_evaluation.html#regression-metrics}},
note         = {Accessed 2026-05-10}
}

@misc{sklearn_minmax_scaler,
author       = {{scikit-learn developers}},
title        = {{MinMaxScaler}},
year         = {2026},
howpublished = {\url{https://scikit-learn.org/stable/modules/generated/sklearn.preprocessing.MinMaxScaler.html}},
note         = {Accessed 2026-05-10}
}

@inproceedings{fong2018finite,
title={Finite element method solution uncertainty, asymptotic solution, and a new approach to accuracy assessment},
author={Fong, Jeffrey T and Marcal, Pedro V and Rainsberger, Robert and Ma, Li and Heckert, N Alan and Filliben, James J},
booktitle={Verification and Validation},
volume={40795},
pages={V001T12A001},
year={2018},
organization={American Society of Mechanical Engineers}
}

@misc{Yaker2026InverseSRF,
author        = {Yaker, Joseph and Markovic, Jovan and Reineri, Alessandro and Kurkcuoglu, Doga Murat and Zorzetti, Silvia},
title         = {Neural-network inverse design of {SRF} cavities and transmons for bosonic quantum computation},
year          = {2026},
eprint        = {2607.02289},
archivePrefix = {arXiv},
primaryClass  = {quant-ph},
url           = {https://arxiv.org/abs/2607.02289}
}

@misc{mohseni2024buildquantumsupercomputerscaling,
title={How to Build a Quantum Supercomputer: Scaling from Hundreds to Millions of Qubits}, 
author={Masoud Mohseni and Artur Scherer and K. Grace Johnson and Oded Wertheim and Matthew Otten and Namit Anand and Navid Anjum Aadit and Yuri Alexeev and Gilad Ben-Shach and Kirk M. Bresniker and Kerem Y. Camsari and Barbara Chapman and Soumitra Chatterjee and Shuvro Chowdhury and Gebremedhin A. Dagnew and Tom Dvir and Aniello Esposito and Farah Fahim and Michael Ferguson and Marco Fiorentino and Archit Gajjar and Katerina Gratsea and Gaurav Gyawali and Christian Heiter and Ali H. Z. Kavaki and Abdullah Khalid and Xiangzhou Kong and Bohdan Kulchytskyy and Elica Kyoseva and Ruoyu Li and P. Aaron Lott and Igor L. Markov and Robert F. McDermott and Lucas Morais and Giacomo Pedretti and Pooja Rao and Eleanor Rieffel and Allyson Silva and John Sorebo and Panagiotis Spentzouris and Ziv Steiner and Boyan Torosov and Davide Venturelli and Robert J. Visser and Zak Webb and Xin Zhan and Yonatan Cohen and Pooya Ronagh and Alan Ho and Raymond G. Beausoleil and John M. Martinis},
year={2026},
note={arXiv:2411.10406v3}, 
eprint={2411.10406},
archivePrefix={arXiv},
primaryClass={quant-ph},
url={https://arxiv.org/abs/2411.10406}, 
}

@article{Stuart_2010, title={Inverse problems: A Bayesian perspective}, volume={19}, DOI={10.1017/S0962492910000061}, journal={Acta Numerica}, author={Stuart, A. M.}, year={2010}, pages={451–559}}

@article{Cotter_2009,
doi = {10.1088/0266-5611/25/11/115008},
url = {https://doi.org/10.1088/0266-5611/25/11/115008},
year = {2009},
month = {oct},
publisher = {},
volume = {25},
number = {11},
pages = {115008},
author = {Cotter, S L and Dashti, M and Robinson, J C and Stuart, A M},
title = {Bayesian inverse problems for functions and applications to fluid mechanics},
journal = {Inverse Problems}}

@misc{taghizadeh2024bayesianinversionidentificationdoping,
      title={Bayesian inversion for the identification of the doping profile in unipolar semiconductor devices}, 
      author={Leila Taghizadeh and Ansgar Jüngel},
      year={2024},
      eprint={2408.11485},
      archivePrefix={arXiv},
      primaryClass={math.NA},
      url={https://arxiv.org/abs/2408.11485}, 
}

@article{hornik1989multilayer,
  title   = {Multilayer feedforward networks are universal approximators},
  author  = {Hornik, Kurt and Stinchcombe, Maxwell and White, Halbert},
  journal = {Neural Networks},
  volume  = {2},
  number  = {5},
  pages   = {359--366},
  year    = {1989},
  doi     = {10.1016/0893-6080(89)90020-8}
}

@misc{omalley2019kerastuner,
  title        = {KerasTuner},
  author       = {O'Malley, Tom and Bursztein, Elie and Long, James and Chollet, Fran\c{c}ois and Jin, Haifeng and Invernizzi, Luca and others},
  year         = {2019},
  howpublished = {\url{https://github.com/keras-team/keras-tuner}}
}

\end{document}